\documentclass[apj]{emulateapj}
\newcommand{\beq}{\begin{equation}}
\newcommand{\eeq}{\end{equation}}
\newcommand{\beqn}{\begin{eqnarray}}
\newcommand{\eeqn}{\end{eqnarray}}
\usepackage{color}
\begin{document}

\author{Brian D.~Farris\altaffilmark{1,2}, Paul~Duffell\altaffilmark{1}, Andrew I.~MacFadyen\altaffilmark{1}, and Zoltan~Haiman\altaffilmark{2}} 
\altaffiltext{1}{Center for Cosmology and Particle Physics, New York University}
\altaffiltext{2}{Department of Astrophysics, Columbia University}

\title{binary black hole accretion from a circumbinary disk: gas dynamics inside the central cavity}

\begin{abstract}
We present the results of 2D hydrodynamical simulations of circumbinary disk accretion using the finite-volume code {\it DISCO}. This code solves the 2D viscous Navier-Stokes equations on a high-resolution moving mesh which shears with the fluid flow, greatly reducing advection errors in comparison with a fixed grid. We perform a series of simulations for binary mass ratios in the range $0.026 \le q \le 1.0$, each lasting longer than a viscous time so that we reach a quasi-steady accretion state. In each case, we find that gas is efficiently stripped from the inner edge of the circumbinary disk and enters the cavity along accretion streams, which feed persistent ``mini-disks" surrounding each black hole. We find that for $q \gtrsim 0.1$, the binary excites eccentricity in the inner region of the circumbinary disk, creating an overdense lump which gives rise to enhanced periodicity in the accretion rate. The dependence of the periodicity on mass ratio may provide a method for observationally inferring mass ratios from measurements of the accretion rate. We also find that for all mass ratios studied, the magnitude of the accretion onto the secondary is sufficient to drive the binary toward larger mass ratio. This suggests a mechanism for biasing mass ratio distributions toward equal mass.
\end{abstract}

\section{Introduction}
It is currently believed that supermassive black holes (SMBHs) with masses between $10^4$ and $10^9 M_{\sun}$ reside in nearly all nearby galaxy nuclei \citep{kormendy95,ferrarese05}. Furthermore, galaxy mergers are believed to give rise to bound SMBH binary systems in the merged galaxy remnant, following a phase of dynamical friction in which the SMBHs become gravitationally bound \citep{begelman80,roos81,merritt05}. Such galaxy mergers occur commonly in models of hierarchical structure formation \citep{haehnelt02}. 

It is expected that there is an abundance of dense gas in the nuclei of merged galaxies \citep{barnes92,springel05} and that this gas can form a circumbinary accretion disk \citep{artymowicz96,armitage02,milos05}. In this scenario, tidal torques from the binary tend to drive gas outward, clearing an empty hollow around the binary. Viscous torques transport angular momentum outward in the disk, allowing gas to flow inward and refill the cavity. The balance of tidal and viscous torques determines the location of the inner edge of the circumbinary disk at $r\approx 2 a$, where $a$ is the binary separation. This balance can be maintained, provided the timescale $t_{gw}$ for inspiral of the binary due to gravitational wave emission is much longer than the viscous timescale of the disk, $t_{visc}$. This is known as the ``pre-decoupling" epoch, and is the focus of this
paper. Such binaries may provide a unique opportunity to observe
various time-dependent and kinematic electromagnetic signatures
originating from the interaction of the binary with the surrounding
accretion disk.  The gravitational radiation originating from the SMBH
binary inspiral should also be detectable by Pulsar Timing Arrays
(PTAs) for massive ($\sim 10^8-10^9 {\rm M_\odot}$) binaries at
redshift $z\approx 1$ - typically these sources are around the time
of decoupling \citep{tanaka12,lommen12,sesana12}. Typical binary BH GW
sources detected by a space interferometer at higher frequencies, such
as eLISA \citep{amaroseoane13}, would be well past the decoupling epoch, but the
lowest-mass, unequal-mass eLISA sources may still be surrounded by a
circumbinary disk \citep{kocsis11, kocsis12b}.

Aspects of this problem have been studied analytically \citep{goldreich80,artymowicz94,ivanov99,armitage02,milos05,chang10,haiman09,shapiro10,liu10,kocsis12a,kocsis12b,tanaka13,shapiro13}, often using approximate angle-averaged tidal torque formulae. While this technique has proven very useful in highlighting qualitative features of the accretion, 2D or 3D simulations are required to capture the effect of gas streams which may be stripped off the inner cavity, accreting directly onto the binary. Prior numerical simulations have been performed in 2D \citep{macfadyen08} (hereafter MM08), \citep{cuadra09,dorazio13} and in 3D \citep{hayasaki07,farris11,farris12,noble12,shi12,roedig12,bode12,giacomazzo12}. However, many simulations to date have excised the innermost region surrounding the binary and imposed an inner boundary condition, potentially neglecting important dynamics occuring inside the excized region. Such simulations are also unable to track the accretion onto each individual black hole. All examples of simulations which capture the inner cavity have either been performed with Smoothed Particle Hydrodynamics (SPH) codes \citep{hayasaki07,cuadra09,nixon11,roedig12}, or have been restricted to relatively small numbers of orbits \citep{giacomazzo12,farris12}. This will be the first 2D study including the inner cavity, using shock-capturing Godunov-type methods evolving  thin ($h/r \sim 0.1$) disks over the viscous timescales necessary to accurately capture the steady-state accretion at high resolution. Such methods have already been shown to be effective for the related problem of gap opening in proto-planetary disks \citep{duffell13}.

Our simulations of viscous disk accretion onto a binary use {\it DISCO}, a high-resolution, moving-mesh finite-volume code \citep{duffell11,duffell12,duffell13} in which we impose no inner boundary condition, so that the black holes are included in the computational domain. Because grid cells move azimuthally with the fluid, we minimize advection errors, allowing us to accurately capture dynamics of accretion streams which penetrate into the cavity. We measure accretion rates onto each black hole and compare results for mass ratios $q \equiv m_2/m_1$ between $0.026$ and $1.0$.

The structure of the paper is as follows. In Sec.~\ref{sec:methods}, we review the viscous-hydrodynamic evolution equations and summarize our numerical methods. In Sec.~\ref{sec:results} we describe our results and compare them with previous work. In Sec.~\ref{sec:conclusions} we summarize our conclusions and outline topics for future work. Throughout the paper, we adopt units in which $G=c=1$.

\section{Methods}
\label{sec:methods}
All simulations are performed using the {\it DISCO} code \citep{duffell12,duffell13}. In regions where the Keplerian motion dominates, the method is effectively Lagrangian, and advection errors are greatly reduced. {\it DISCO} is based on the {\it TESS} code \citep{duffell11}, and shares many of its features. For more details on the methods employed by {\it DISCO} and {\it TESS}, see \citet{duffell11,duffell12}.
\subsection{Basic Equations}

We model the flow of a thin accretion disk onto a SMBH binary using the vertically averaged, 2D Navier-Stokes equations,
\begin{equation}
  \partial_t \Sigma + \frac{1}{r} \partial_r \left( r \Sigma v_r \right) +  \frac{1}{r} \partial_{\phi} \left(\Sigma v_{\phi} \right)  = 0
\end{equation}

\begin{eqnarray}
  \partial_t \left( \Sigma v_r \right) &+& \frac{1}{r} \partial_r \left( r\left(\Sigma v_r^2 + P\right)\right) + \frac{1}{r} \partial_{\phi} \left( \Sigma v_r v_{\phi}\right)\nonumber\\ &=& (\Sigma v_{\phi}^2 + P)/r + F^{vis}_r + F^{grav}_r
\end{eqnarray}

\begin{eqnarray}
  \partial_t \left(r \Sigma v_{\phi}\right)&+& \frac{1}{r} \partial_r \left(r^2 (\Sigma v_r v_{\phi} ) \right) +  \frac{1}{r} \partial_{\phi} \left(r(\Sigma v_{\phi}^2 + P ) \right)\nonumber\\&=& F^{vis}_{\phi} + F^{grav}_{\phi} \ .
\end{eqnarray}

Here $\Sigma$ is the surface density, $P$ is the pressure, and $v_r$ and $v_{\phi}$ are the radial and azimuthal fluid velocities, respectively. $F^{vis}_{\phi}$ and $F^{grav}_{\phi}$ are the viscous and gravitational source terms, which are described in Sec.~\ref{sec:visc} and Sec.~\ref{sec:grav}. All quantities should be interpreted as vertically averaged. The pressure is set to enforce a locally isothermal equation of state of the form, 
\begin{eqnarray}
  c_s = \left(
  \frac{P}{\Sigma}
\right)^{1/2} &=&   (h/r) 
\left(
  \frac{M_1}{r_1} + \frac{M_2}{r_2} 
\right)^{1/2}\nonumber\\
&\approx&  
\left\{ \begin{array}{cc}
    (h/r) (r_1 \Omega_{k,1}) \ \ \ \ \ & r_1 \ll r_2 \\
    (h/r) (r_2 \Omega_{k,2}) \ \ \ \ \ & r_2 \ll r_1\\
    (h/r) (r \Omega_{k}) \ \ \ \ \ \ & r \gg a \ .
  \end{array}
  \right.
\end{eqnarray}
 Here $r_1$ and $r_2$ are the distances from the black holes labelled $bh_1$ and $bh_2$ respectively, $\Omega_{k,1}$ and $\Omega_{k,2}$ are the corresponding keplerian angular velocites, $r$ is the distance from the origin, and $\Omega_k$ is the Keplerian orbital velocity around a central point mass of mass $M = m_1+m_2$, and $a$ is the binary separation. Following \citet{macfadyen08,dorazio13}, we choose $h/r = \mbox{const} = 0.1$. This choice of temperature profile matches that of a constant $h/r=0.1$ $\alpha-$disk near each individual black hole in the limit $r_{1,2} \ll r$, and that of a constant $h/r=0.1$ circumbinary $\alpha-$disk in the limit that $r \gg r_{1,2}$. We note that $\alpha$-disks may be colder and have $h/r \sim 10^{-3}$.
However, simulating disks numerically with $h/r=10^{-3}$ is currently prohibitively challenging, and our choice of $h/r=0.1$ allows for comparison with previous work \citep{artymowicz94,macfadyen08,dorazio13}. In future work, we intend to investigate changes which emerge at smaller $h/r$, as well as relaxing the assumption that $h/r$ is constant everywhere, as the gas in each circumstellar mini-disk may in fact be hotter than the circumbinary gas.
 
\subsection{Viscosity Prescription}
\label{sec:visc}
We choose an $\alpha-$law viscosity of the form \citep{shakura73},
\begin{equation}
  \label{eq:visc_nu}
  \nu = \alpha c_s^2 \Omega_k^{-1} \ ,
\end{equation}
where $\alpha = 0.1$ is chosen for this study. Close to the binary, we expect that as the flow develops a strong radial component in the streams, the $\alpha$-viscosity prescription becomes inapplicable. In fact, there is some evidence from 3D MHD simulations of accretion onto binaries that the effective $\alpha$ may increase significantly near the inner edge of the disk and inside the cavity \citep{shi12,noble12}. An analagous effect is also associated with the position of the innermost stable circular orbit in fully relativistic calculations \citep{penna13}. However, the solution should not be sensitive to the details of the viscosity prescription inside the cavity, as  the flow in this region is dominated by variations occuring on the orbital timescale, which is much shorter than the viscous timescale. Thus, viscosity is not expected to play an important role until the gas settles into the ``mini-disks" around each individual black hole (BH). We therefore treat the role of viscosity in driving accretion of the mini-disks with an accretion prescription (see Sec.~\ref{sec:accretion}).

The presence of viscosity modifies the hydrodynamic equations by adding a viscous component to the stress tensor given by
\begin{equation}
  \sigma_{ij} = \Sigma \nu \left(
    \nabla_iv_j + \nabla_j v_i - \delta_{ij}\nabla \cdot v
  \right)
\end{equation}
where we have set the bulk viscosity to $0$. Expressing components of the divergence of the viscous stress tensor in cylindrical coordinates, we find (see Appendix A of \citealt{dorazio13}),
\begin{eqnarray}
  F^{vis}_r &\equiv& (\nabla \cdot \sigma)_r= \frac{1}{r^2} \partial_r \left( r \Sigma \nu 
\left( r \partial_r v_{\hat{r}} - v_{\hat{r}} + \Omega r \frac{\partial_{\phi}P}{P}\right) \right) \nonumber\\
&&+ \frac{1}{r^2}\partial_{\phi}\left(
\Sigma \nu 
\left( \partial_{\phi} v_{\hat{r}} - \Omega r^2 \left(\frac{7}{2} \frac{1}{r} + \frac{\partial_{r}P}{P}\right) \right) \right) 
               \end{eqnarray}

               \begin{eqnarray}
                 F^{vis}_{\phi} &\equiv& (\nabla \cdot \sigma)_{\phi} = \frac{1}{r} \partial_r \left(r \Sigma \nu 
               \left(r^2 \partial_r \Omega - v_{\hat{r}} \frac{\partial_{\phi}P}{P} \right)\right)\nonumber\\
               &&+  \frac{1}{r}\partial_{\phi}\left(
               \Sigma \nu 
               \left(r \partial_{\phi} \Omega + r v_{\hat{r}}  \left(\frac{7}{2} \frac{1}{r} + \frac{\partial_{r}P}{P}\right) \right)
             \right)
           \end{eqnarray}
           where we use gradients of fluid velocity and pressure which have been computed for the reconstruction step \citep{duffell11}. We have used the following relations, which can be derived from Eq.~\ref{eq:visc_nu},
           \begin{eqnarray}
             \frac{\partial_{r}(r^2 \Sigma \nu)}{r^2 \Sigma \nu} &=& \frac{7}{2} \frac{1}{r} + \frac{\partial_{r}P}{P}\\
             \frac{\partial_{\phi}(r^2 \Sigma \nu)}{r^2 \Sigma \nu} &=& \frac{\partial_{\phi}P}{P}
           \end{eqnarray}

           \subsection{Gravity}
\label{sec:grav}
           
           The gravitational source terms are given by:
           \begin{eqnarray}
             F^{grav}_r &=& -\frac{\partial \Phi}{\partial r}\\
             F^{grav}_{\phi} &=&  -\frac{1}{r}\frac{\partial \Phi}{\partial \phi} \ ,
           \end{eqnarray}
           where the $\Phi$ is the Newtonian gravitational potential due to the binary,
           \begin{eqnarray}
             \Phi(r,\phi) &=& - \frac{M_1 }{\left(\left|\mathbf{r}-\mathbf{r}_{bh1}\right|^2 + \epsilon_s^2 \right)^{1/2}}\nonumber\\
                          &&- \frac{M_2 }{\left(\left|\mathbf{r}-\mathbf{r}_{bh2}\right|^2 + \epsilon_s^2 \right)^{1/2}} \ .
           \end{eqnarray}
           Here, we have followed \citet{duffell13,tanaka02,masset02,muller12} and used a ``vertically averaged" potential which attempts to account for corrections to the gravitational force near each point mass due to the nonzero thickness of the disk. This choice of potential also conveniently avoids divergences at the location of each point mass. In each of our simulations,  we choose a smoothing length $\epsilon_s / a = 0.05$.
           \subsection{Accretion Prescription}
           \label{sec:accretion}
           Inside the cavity, matter accumulates near each BH and forms ``mini-disks" of gas orbiting the individual BHs. In order to achieve a quasisteady state, we must account for the draining of these ``mini-disks" due to accretion onto each BH. In the absense of a fully relativistic code in which the flow accross the event horizon can be resolved, this necessitates an accretion prescription. We treat each mini-disk as an alpha-disk with $h/r = 0.1$ and $\alpha=0.1$, and compute the viscous timescale at cells near each BH. The viscous timescale for $bh_i$ is,
           \begin{eqnarray}
           \label{eq:tvis}
             t_{vis,i} &=& \frac{2}{3} \frac{r_i^2}{\nu_i} \nonumber\\
                                 &=& \frac{1}{3 \pi} \frac{1}{\alpha (h/r)^2} \left(\frac{r_i}{a}\right)^{3/2} \mu_i^{-1/2}t_{bin}
           \end{eqnarray}
           where $r_i$ is the distance from $bh_i$, $\mu_i \equiv m_i / M$ is the ratio of the mass of $bh_i$ to the total binary mass, and $t_{bin}$ is the binary orbital period. When $r_i/a < 0.5$  we add a source term to the right hand side of the continuity equation of the form,
           \begin{equation}
           \label{eq:rhosink}
             \left(\frac{d \Sigma}{dt}\right)_{sink,i} = - \frac{\Sigma}{t_{vis,i}}
           \end{equation}
           This removes fluid from the vicinity of each BH at the rate expected according to the $\alpha-$disk model.  We have experimented with varying the radius within which we apply the accretion prescription and find that the results are not sensitive to this parameter. This is expected, as $d \Sigma / dt \propto \Sigma r_i^{-3/2}$, and $\Sigma$ itself is peaked near each BH. Thus, the majority of the fluid which is removed comes from very close the BHs.

           \subsection{Initial Setup}
           Following MM08, we use an initial density profile of the form
           \begin{equation}
             \Sigma(r,t_0) = \Sigma_0 \left(\frac{r}{r_s}\right)^{-3} \mbox{exp}\left[ - \left( \frac{r}{r_s}\right)^{-2}\right]
           \end{equation}
           and a near-Keplerian velocity profile, modified to account for binary torques and pressure gradients.,
           \begin{equation}
             \Omega^2 \approx \Omega_k^2\left(1+\frac{3}{16}(r/a)^{-2}\right)^2 + \frac{1}{r\Sigma}\frac{dP}{dr}
           \end{equation}
           We also add radial velocity to account for viscous drift,
           \begin{equation}
             v_r = \frac{2}{r^2 \Omega \Sigma} \frac{\partial(r^2 \sigma_{r \phi})}{\partial r} \ .
           \end{equation}
           Note that our initial setup exactly mirrors that of MM08, except that we choose $\alpha = 0.1$, instead of $\alpha = 0.01$. The main effect of this choice is to shorten the viscous timescale by a factor of $10$. However, as the viscous timescale is still many times the orbital period, qualitative features of the flow are expected to remain insensitive to the choice of $\alpha$. 

\begin{figure}
\epsscale{1.0}
\plotone{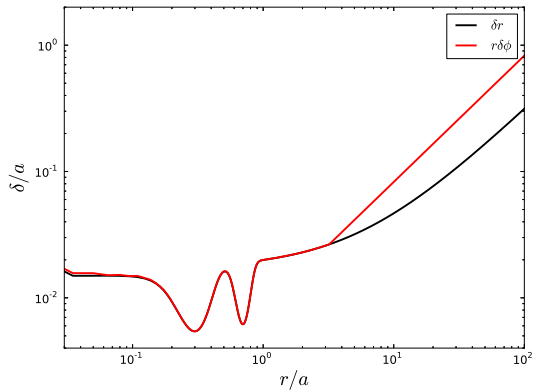}
\caption{Grid cell sizes $\delta r$ (black), and $r \delta \phi$ (red) vs. radius for the $q=0.43$ case. Grid spacing increases roughly logarithmically with radius, and resolution is further increased in annuli near each BH, located at the two sharp minima near $r/a=0.3$ and $r/a=0.7$. }
\label{fig:grid_info}
\end{figure}

For each of our simulations, we exploit the flexibility of the {\it DISCO} code in choosing the grid structure. Outside the cavity, the radial grid spacing increases approximately logarithmically, allowing us to concentrate computational resources on the flow in the inner regions of the disk. Inside the cavity, we also increase the resolution in the annuli surrounding each black hole, in order to maximize accuracy near each point mass. At all radii, we chose our azimuthal grid spacing such that the aspect ratio of each cell is kept approximately $1$, unless this requires a number of azimuthal grid cells $N_{\phi} > 760$, in which case we cap the number of $\phi$-zones and set $N_{\phi}=760$. We have chosen to cap $N_{phi}$ at 760 zones because this ensures that the aspect ratio is close to unity everywhere inside the cavity, where strongly non-axisymmetric structures form, while allowing for fewer zones outside the cavity. In Figure~\ref{fig:grid_info}, we plot the cell spacing in the $r$ and $\phi$ direction vs radius for the $q=0.43$ case in order to illustrate the grid setup.

\section{Results}
\label{sec:results}
    \subsection{Shear Viscosity Test}
\begin{figure}
          \epsscale{1.0}
          \plotone{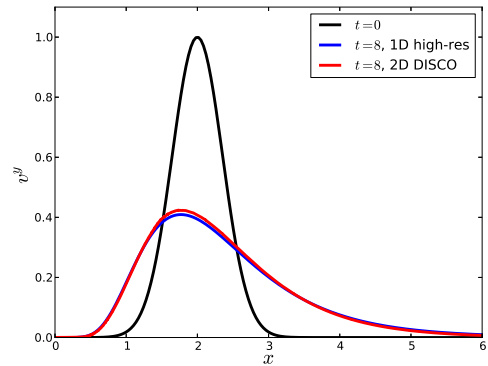}
          \caption{Results of cartesian shear test. The black curve denotes the initial velocity profile at $t=0 $, the blue curve denotes the semi-analytic solution obtained from a high-resolution 1D solver at $t=8.0$, and the red curve denotes the results of the {\it DISCO} code $t=8.0$. }
          \label{fig:shear}
          \end{figure}

    In order to test our viscosity implementation, we have performed a ``cartesian shear test" similar to the one described in Sec.~A7 of \citet{dorazio13} in which we follow the viscous spreading of a shearing velocity profile. As initial data we choose constant density and pressure
    \begin{eqnarray}
    \rho &=& 1.0\\
    P &=& 0.1
    \end{eqnarray}
    and a velocity profile of the form
    \begin{eqnarray}
    v^x &=& 0\\
          v^y &=& \mbox{exp}\left(-\frac{(x-x_0)^2}{\sigma^2}\right)
          \end{eqnarray}
          where $\sigma=0.5$. The most notable difference between our test and that of \citet{dorazio13} is that we choose a viscosity profile of the form,
          \begin{equation}
          \nu = \nu_0 \frac{\Gamma P}{\rho} x^{2} \ .
          \end{equation}
          This choice was made in order to test the ability of {\it DISCO} to accurately treat a spatially varying viscosity, as is present in an $\alpha-$disk. We perform this test using the cylindrical grid employed by {\it DISCO}, thus testing the balance of all components of the viscous forces and hydrodynamic fluxes and source terms in all directions. We also choose a grid in which the resolution becomes more coarse with radius, as in our binary simulations. Specifically, we set $\delta r \propto r+1$, we use $128$ radial zones, and we keep $\delta \phi \approx \delta r$. Our choice of viscosity no longer allows a simple analytic solution for the time evolution of the fluid velocity, so we instead compare the velocity with the result of a high-resolution 1D code which solves the simplified fluid equation
          \begin{equation}
          \partial_t v_y(x,t) = \partial_x \left( \nu \partial_x v_y \right)\ .
          \end{equation}
          We find excellent agreement between the results of our {\it DISCO} run and that of the 1D code, as shown in Figure~\ref{fig:shear}. In particular our implementation of viscos forces in {\it DISCO} is able to accurately capture the asymmetric spreading caused by a spatially varying viscosity.

\subsection{Gas Dynamics Near the Binary's Orbit}
\begin{deluxetable}{cc}
 \tablecaption{Summary of mass ratios and average accretion rates.
 \label{table:q}}
 \tablewidth{0pt}
 \tablehead{
 \colhead{Mass Ratio $q$} \ \ \ \   &   \colhead{$\langle \dot{M}\rangle / \langle \dot{M}_0 \rangle$ }
} 
 \startdata
$0.026$&1.06\\
$0.053$&1.56\\
$0.11$&1.76\\
$0.25$&1.68\\
$0.43$&1.62\\
$0.67$&1.60\\
$0.82$&1.58\\
$1.0$&1.55\\
 \enddata
\end{deluxetable}
We restrict our attention to binaries with separations $a/M \gg 100$, so that we are in the pre-decoupling epoch in which the gravitational wave inspiral timescale is long compared to the viscous timescale, and orbital shrinkage can be ignored. We also note that the orbital decay due to interaction with the disk will be much slower than the viscous timescale in this regime \citep{kocsis12b}. Thus, we expect the flow to reach a quasi-steady state in which the size of the inner cavity is determined by the balance between gravitational torques and viscous stresses \citep{artymowicz94,milos05}. In order to realize this quasi-steady state over the region of interest in our simulations, we must evolve for at least one viscous timescale at the radius of the cavity wall at $r/a \sim 2$.
\begin{equation}
t_{vis}(r) \sim \frac{2}{3} \left[\alpha \left(\frac{h}{r}\right)^2 \Omega(r)\right]^{-1} \sim 300 \left(\frac{r}{2a}\right)^{3/2} t_{bin}
\end{equation}
In each of our simulations, we evolve for $\sim 1.5 t_{vis}(2a)$. We find that this is sufficient in order to reach a quasi-steady state as reflected in relatively steady density profiles that we achieve after $t\gtrsim t_{vis}$. The mass ratios and time averaged accretion rates for each simulation are summarized in Table~\ref{table:q}. Time average accretion rates are normalized by the time averaged accretion rate onto a single BH, $\dot{M}_0$. We note that although the normalized accretion rate tends toward unity for small $q$ as expected, it remains greater than unity for all cases considered. This is somewhat surprising, as the clearing of a cavity by binary torques has been proposed as a mechanism for surpressing the accretion rate onto the binary \citep{milos05}. However, such arguments may underestimate the role that non-axisymmetric accretion streams play in allowing gas to penetrate into the cavity. Furthermore, binary torques may be responsible for moving gas near the inner disk edge onto more eccentric orbits, causing them to be captured by one of the BHs, thus increasing the accretion rate relative to that of a single BH. 

\begin{figure*}
\epsscale{0.25}
\plotone{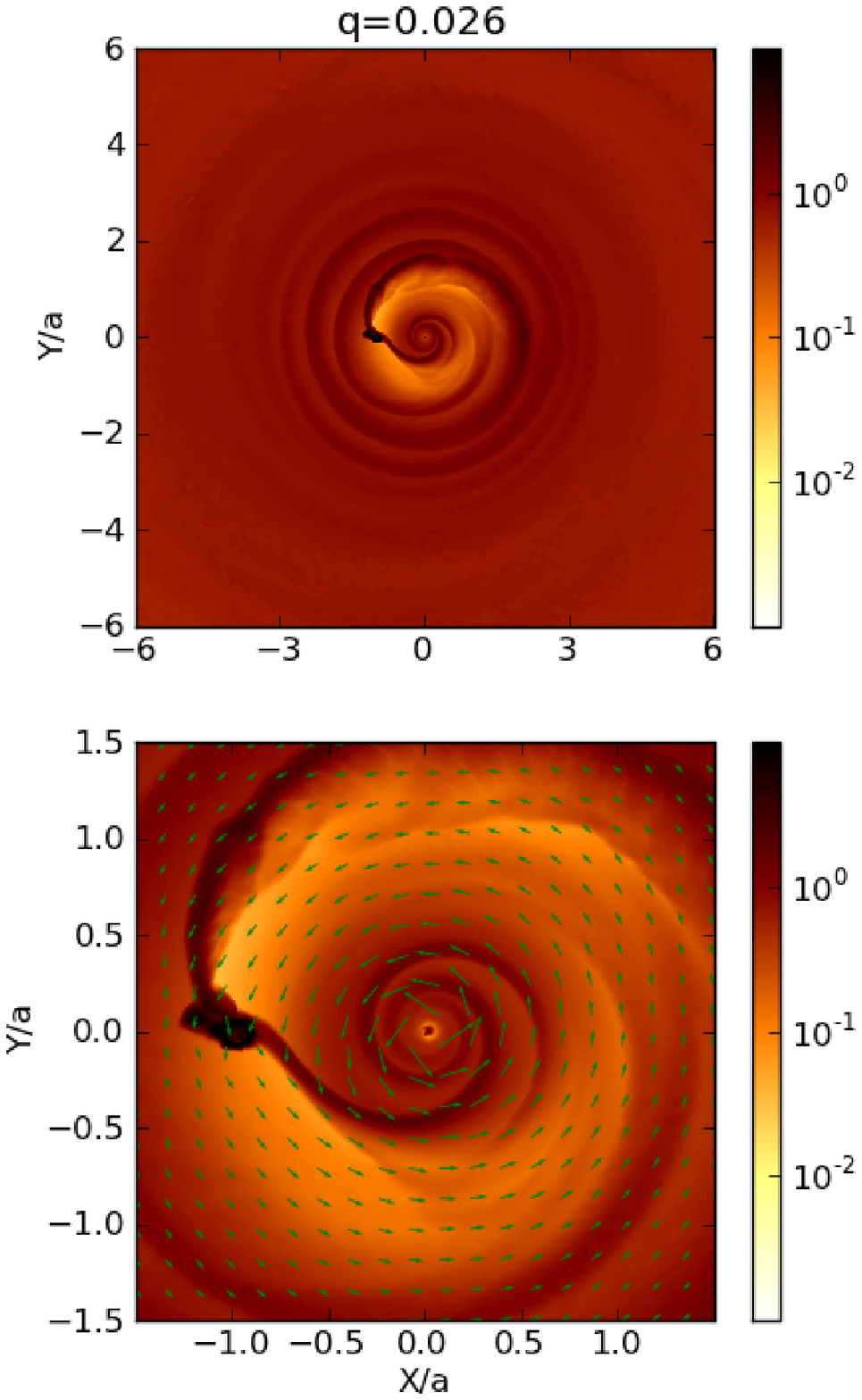}
\plotone{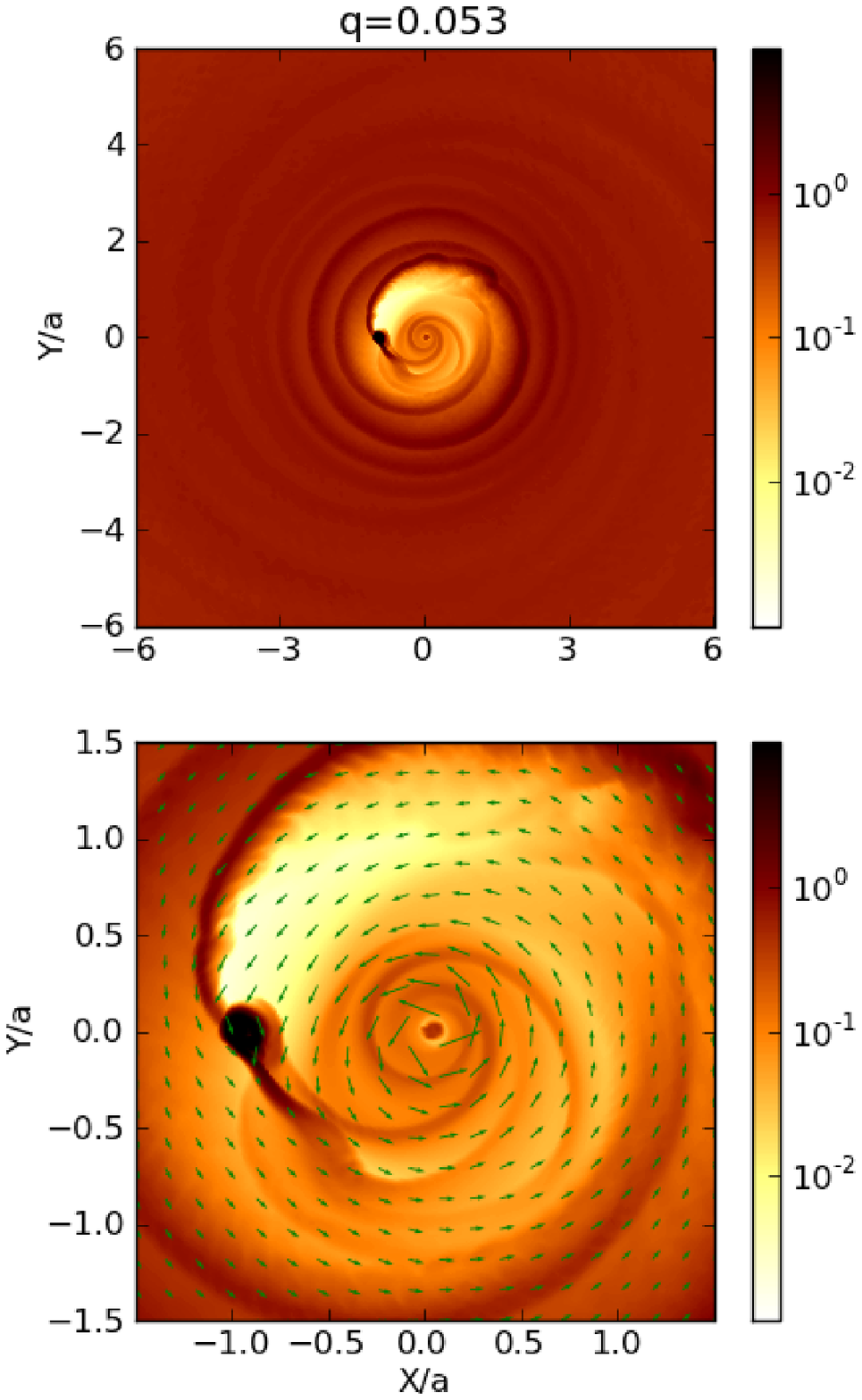}
\plotone{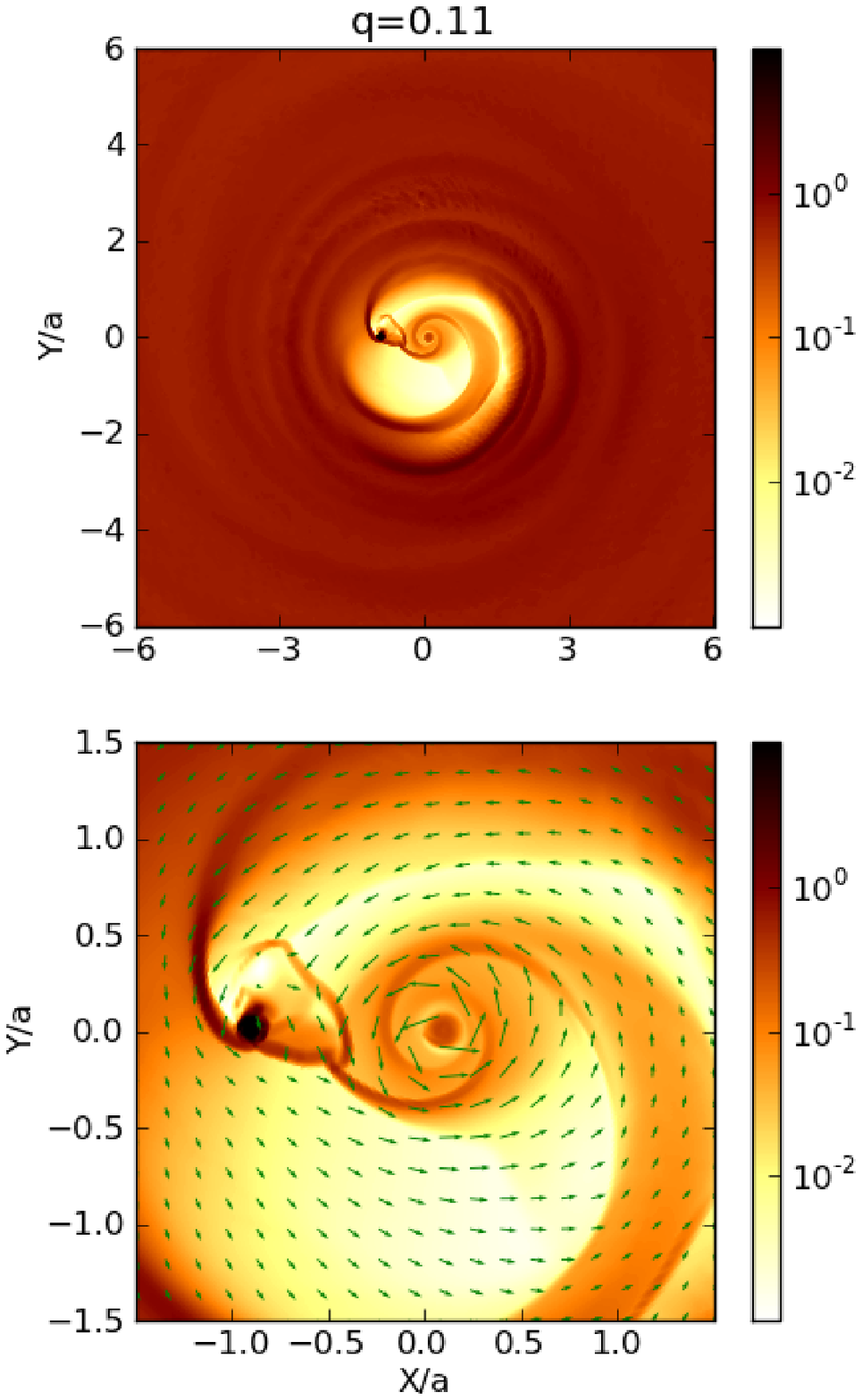}
\plotone{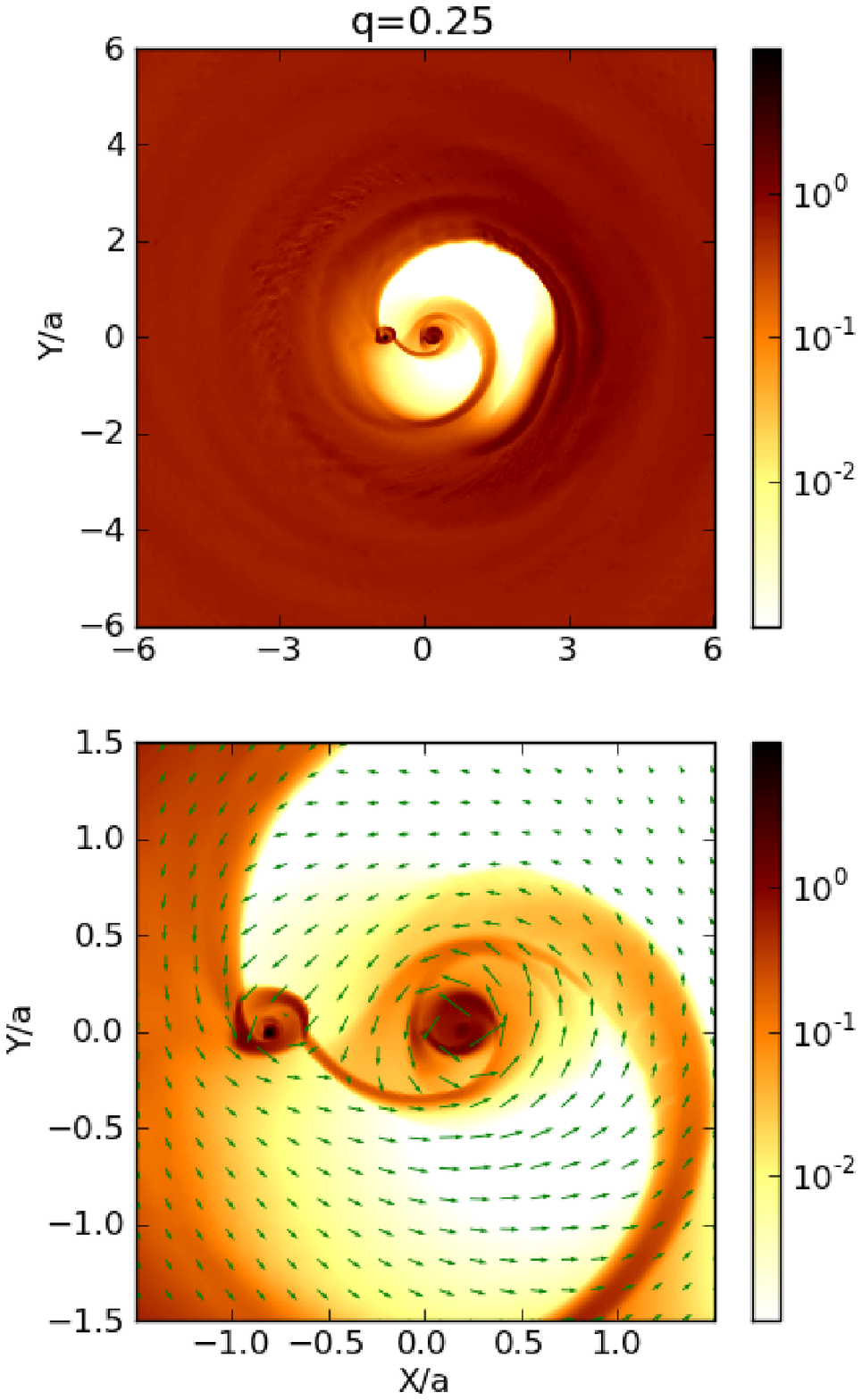}
\plotone{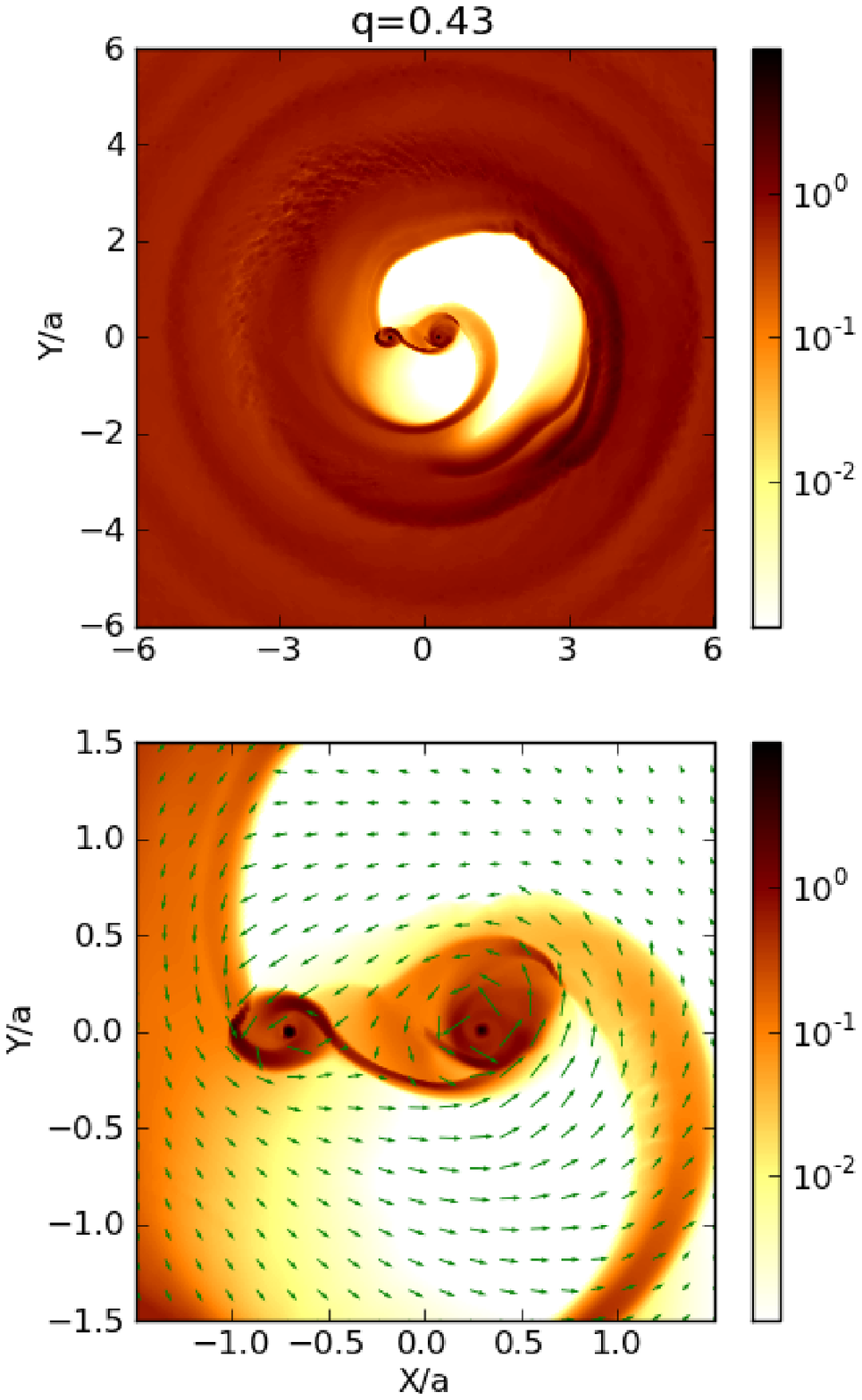}
\plotone{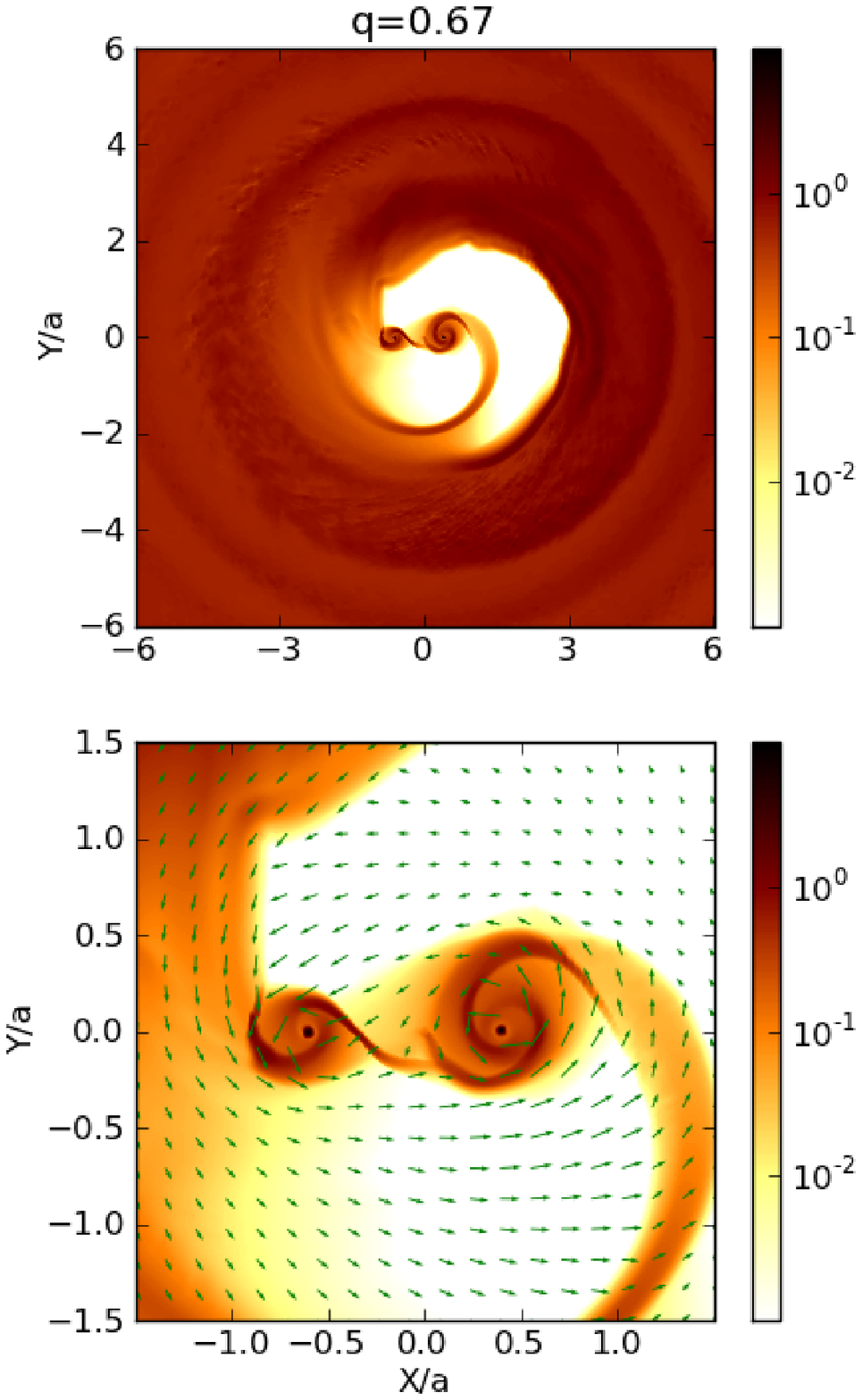}
\plotone{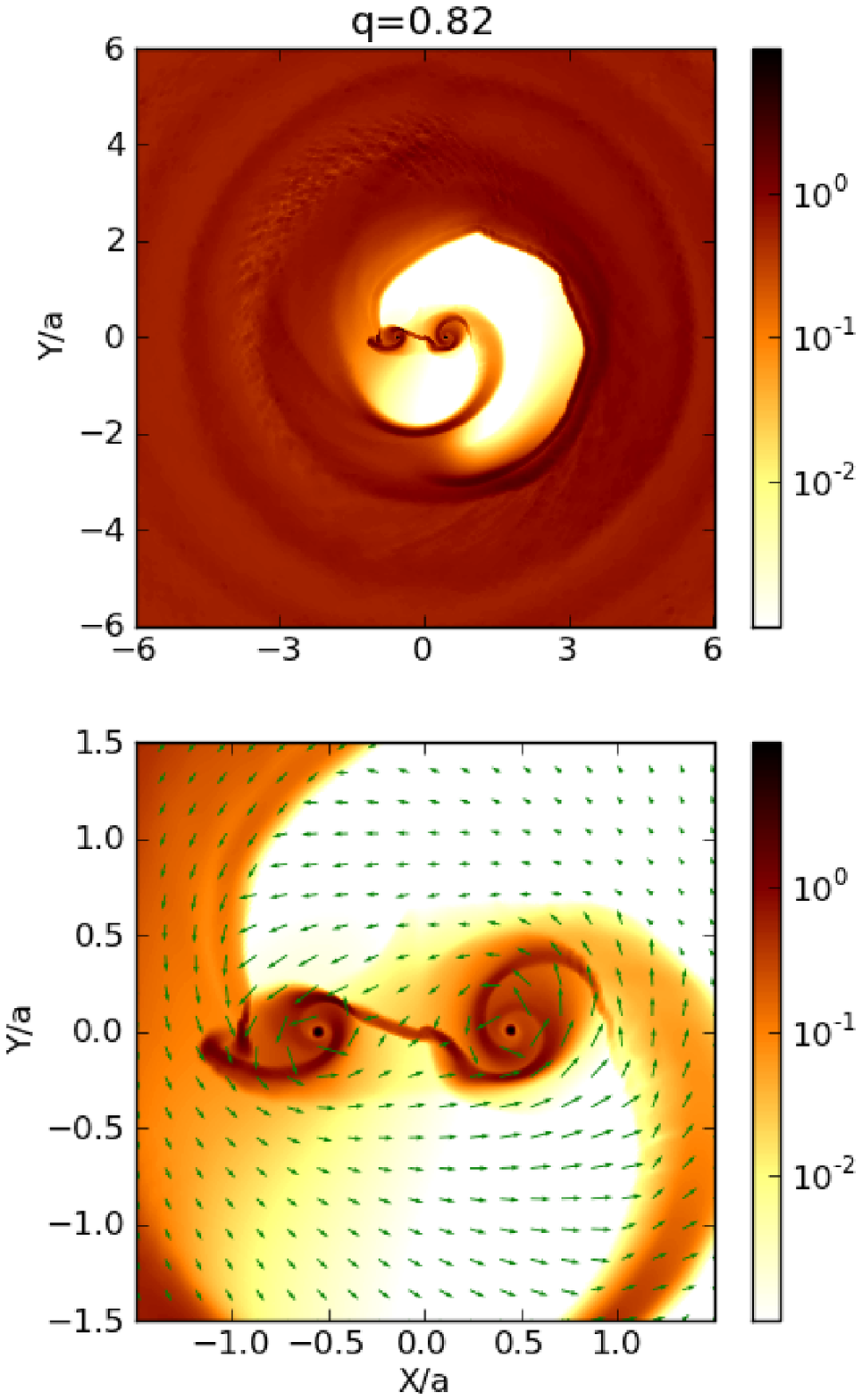}
\plotone{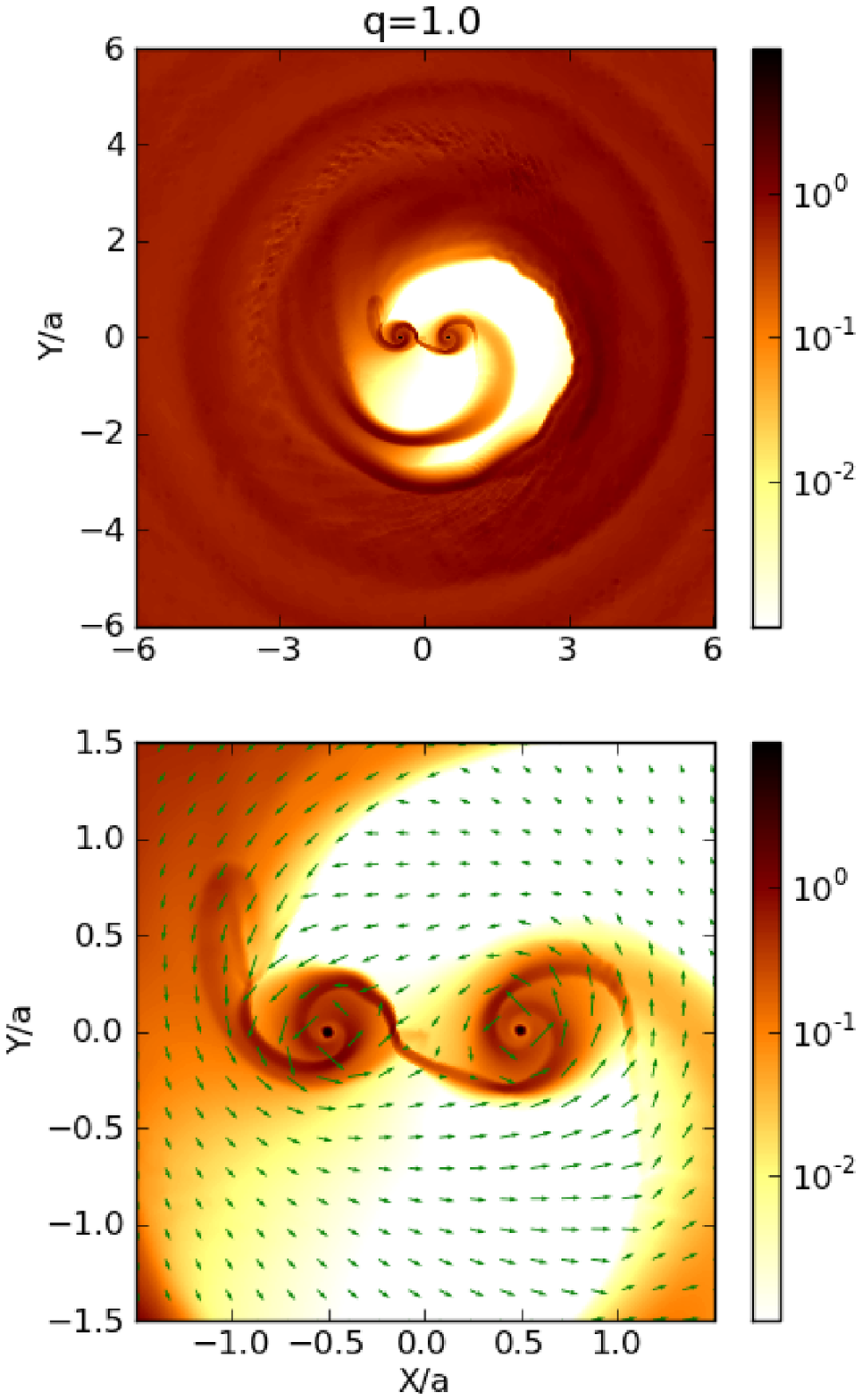}
\caption{Snapshots of surface density $\Sigma$ during quasi-steady state after $t \approx 460 t_{bin} \approx 1.5 t_{vis}$. Surface density is normalized by the maximum value at $t=0$ and plotted on a logarithmic scale. For each snapshot, we plot both the inner $\pm 6a$ (top panel in each pair), and the inner $\pm 1.5 a$ (bottom panel in each pair). Mass-ratios are, from left to right and top to bottom, $q=0.026$, $0.053$, $0.11$, $0.25$, $0.43$, $0.67$, $0.82$, and $1.0$. Orbital motion is in the counter-clockwise direction. Green arrows represent fluid velocity.}
\label{fig:dens2d}
\end{figure*}

\begin{figure}
\epsscale{1.0}
\plotone{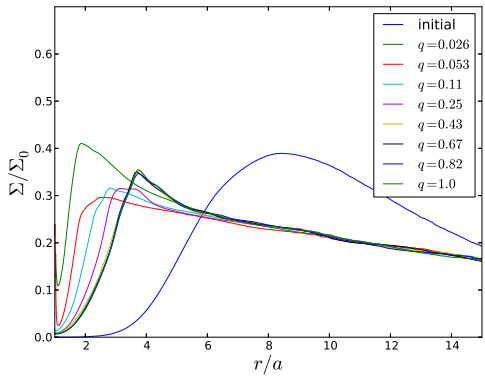}
\caption{Time-averaged, angle averaged surface density as a function of radius for various binary mass ratios. The blue curve is the initial surface density profile. Time average is taken over $446 < t/t_{bin} < 462$. We see that the size of the cavity increases with $q$, but approaches a near-constant profile for $q \gtrsim 0.43$.
}
\label{fig:dens1d}
\end{figure}

\begin{figure}
\epsscale{1.0}
\plotone{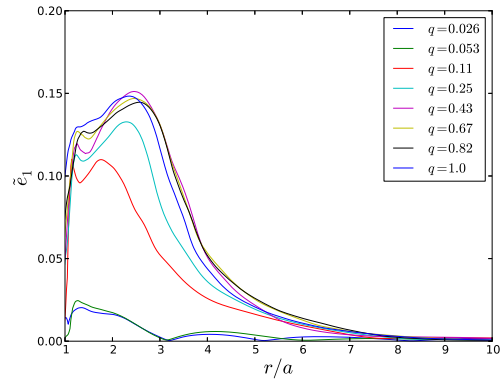}
\caption{Time-averaged eccentricity $\tilde{e}_1$ as a function of radius for various binary mass ratios. Time average is taken over $446 < t/t_{bin} < 462$. Eccentricity is seen to increases with $q$.}
\label{fig:eccentricity}
\end{figure}

Snapshots from each simulation are shown in Figure~\ref{fig:dens2d}. In each case, a low density cavity is maintained surrounding the binary, with the size of the cavity depending on the mass ratio $q$. We find that for $q\lesssim 0.25$, the cavity increases in size and eccentricity with mass ratio $q$. For $q\gtrsim 0.43$, both the size of the cavity and its eccentricity remain roughly constant.  These trends are more obvious in Figure~\ref{fig:dens1d}, where we plot time-averaged, angle-averaged 1D density profiles, and Figure~\ref{fig:eccentricity} where we plot the time-averaged eccentricity profiles. Here, we have used the definition of disk eccentricity provided by MM08, where we take the time-average of the density-weighted m=1 component of the radial velocity, normalized by the density-weighted azimuthal velocity,
\begin{equation}
\tilde{e}_m \sim \frac{| \langle \Sigma v_r e^{im\phi}\rangle|}{\langle\Sigma v_{\phi}\rangle} \ .
\end{equation}
where, 
\begin{equation}
  \langle \cdot \rangle \equiv \frac{1}{2\pi}\int_0^{2 \pi} d \phi \ .
\end{equation}

\begin{figure}
\epsscale{1.0}
\plotone{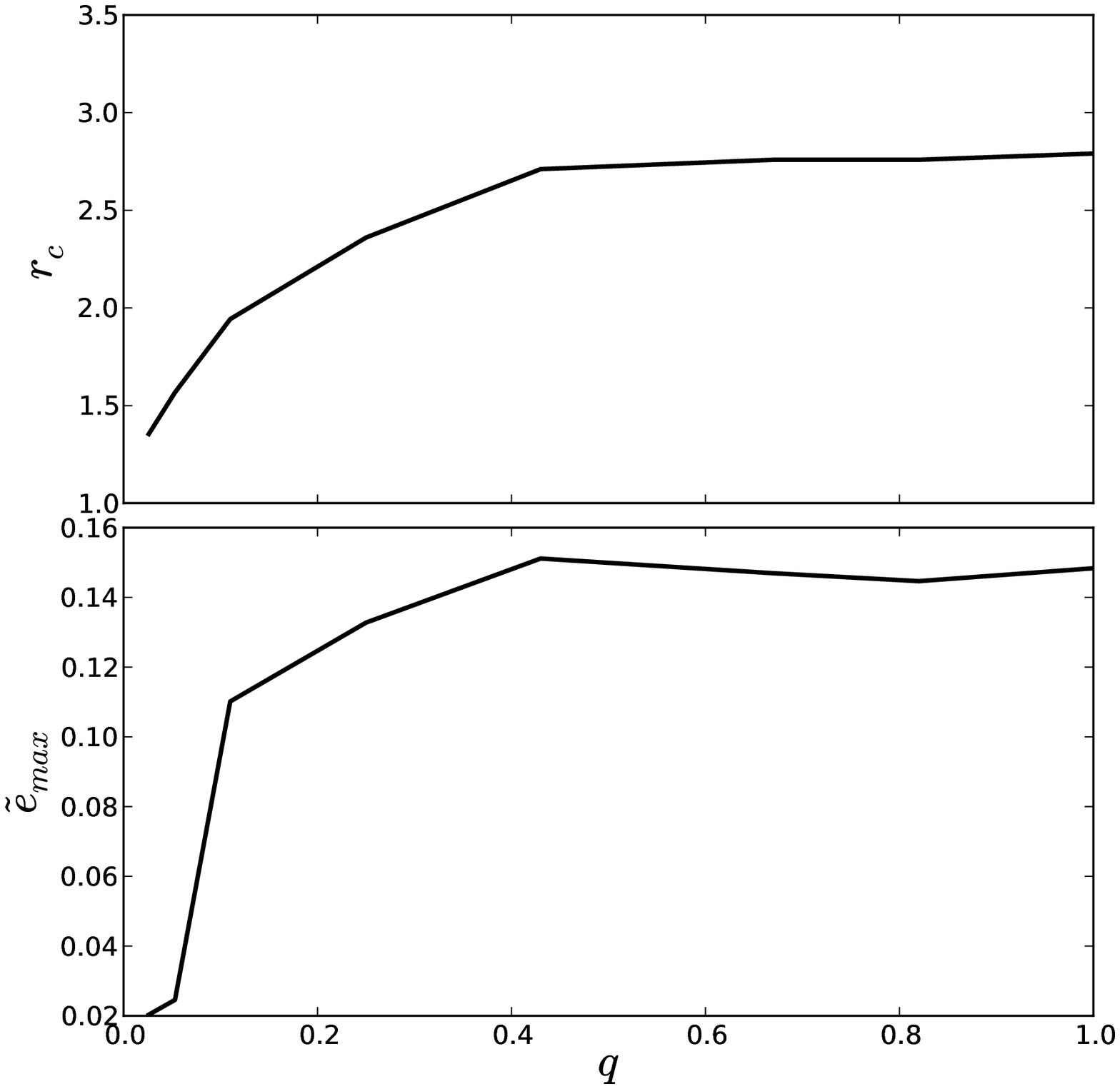}
\caption{Maximum eccentricity $\tilde{e}_{max}$ (bottom) and position of the outer edge of the cavity $r_c$ (top), as a function of mass ratio $q$.}
\label{fig:vs_q}
\end{figure}

We also plot the position of the outer edge of the cavity $r_c$ (defined as the radius at which the $\Sigma=1/2 \Sigma_{max}$), and the maximum eccentricy $\tilde{e}_{max}$ as a function of mass ratio $q$ in Figure~\ref{fig:vs_q}. 

The trend toward an eccentric inner cavity for large mass ratio binaries has now been observed in a number of calculations \citep{macfadyen08,dorazio13,shi12,noble12,farris12}. A plausible explanation for the growth of this eccentricity has been proposed by \citet{dorazio13} and is summarized as follows. Torques from the binary create two streams which enter the cavity. A fraction of the gas in each stream does not accrete onto a BH, but rather is flung outward, impacting the cavity wall on the opposite side from which it entered. If one stream is perturbed and becomes larger, it will push the cavity on the opposite side outward, thus weakening the opposite stream. In this way, the imbalance grows and the inner cavity becomes eccentric. As we have focused on the limit in which $M_{disk} \ll M$ in this study, this disk eccentricity has no effect on the black hole orbit. However, in future work we intend to relax this assumption, as this effect may indeed provide a mechanism for the growth of significant binary eccentricity, which may have important implications for gravitational wave observations. 

We also note that the cavity is not truly evacuated, as substantial gas enters along streams and creates ``mini-disks" surrounding each black hole.  This can be seen in Figure~\ref{fig:dens2d}, in which we plot snapshots of the surface density. We note that in many cases, the local surface density in these ``mini-disks" can exceed that of the circumbinary disk. It has been proposed that such mini-disks can be observed as quasiperiodic AGN sources \citep{artymowicz96,hayasaki08,hayasaki12}. If the accretion timescale of a mini-disk is short compared to the orbital timescale of the binary, each minidisk episode can mimic a stellar tidal disruption event (TDE), and, in case only a single episode is detected, can even be confused with such a TDE \citep{tanaka13}. In our simulations, we find that this is not the case, as the mini-disks are relatively persistent. This is expected in light of our accretion prescription given by Eq.~\ref{eq:tvis}. Scaled to an approximate mini-disk size $r_{md} \approx 0.25a$, we find,
\begin{eqnarray}
\label{eq:tvis_md}
t_{vis,md} &=& 42 \left(\frac{\alpha}{0.1}\right)^{-1}
\left(\frac{h/r}{0.1}\right)^{-2}\nonumber\\
&& \times
\left(\frac{r_{md}}{0.25a}\right)^{3/2}
\left(\frac{q}{0.1}\right)^{-1/2}
t_{bin} \ .
\end{eqnarray}
We note, however, that the viscous timescale of each mini-disk is sensitive to $h/r$, which we hold fixed at $h/r=0.1$ in our model. In a more sophisticated treatment of the temperature, it is possible that the mini-disks can become hotter and thicker, thus decreasing $t_{vis}$. If this heating is sufficient so that $h/r \sim 1$ in a minidisk, we see from Eq.~\ref{eq:tvis} that the viscous timescale can become $t_{vis} \sim 0.42 t_{bin}$, and each minidisk will be effectively drained and replenished during each orbit. We also note that analytic work indicates that binary eccentricity tends to reduce the sizes of both circumprimary and circumsecondary mini-disks \citep{artymowicz94}. This may also shorten the viscous timescale of each mini-disk. We intend to address each of these points in future work.

\subsection{Mass Ratio Evolution}
\begin{figure}
\epsscale{1.0}
\plotone{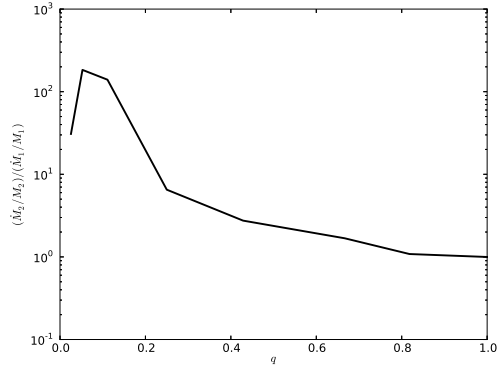}
\caption{Plot of $(\dot{M}_2/M_2)/(\dot{M}_1/M_1)$ vs BH mass ratio $q$. This quantity is $>1$ when $q$ is increasing, and $<1$ when $q$ is decreasing. For each case we find $(\dot{M}_2/M_2)/(\dot{M}_1/M_1) > 1$, indicating that mass ratios are always increasing with time. However, extrapolating the downward turn seen at small $q$, it is possible that there exists a threshold at small $q$, below which $q$ decreases with time.}
\label{fig:mdot_ratio}
\end{figure}

An important question is whether the accretion is causing the mass ratio $q$ to increase or decrease. The growth rate of $q$ is given by,
\begin{equation}
  \frac{dq}{dt} = \frac{d}{dt}\left(\frac{M_2}{M_1}\right) = \frac{M_2}{M_1}\left(\frac{\dot{M}_2}{M_2}-\frac{\dot{M}_1}{M_1}\right) \ ,
\end{equation}
So we see that the condition for $q$ to be increasing is that $\dot{M}_2 / M_2 > \dot{M}_1 / M_1$. In Figure~\ref{fig:mdot_ratio}, we plot the ratio $(\dot{M}_2 / M_2) / (\dot{M}_1 / M_1)$ as a function of $q$. We calculate $\dot{M}_i$ for each BH by computing the total integrated rest mass removed during each timestep according to Eq.~\ref{eq:rhosink}. We find that for all simulations performed, this quantity is greater than unity, indicating that the mass ratio is growing, driving the binary closer to equal mass. For the equal mass case ($q=1.0$), we find that the ratio of accretion rates goes to unity, as expected. This result confirms the findings of a number of previous studies \citep{hayasaki07,cuadra09,roedig11,roedig12,hayasaki12}, all of which found greater accretion onto the secondary. We note, however, that the opposite has been found in simulations of circumbinary accretion onto T Tauri binaries \citep{valborro11}. The fact that $(\dot{M}_2 / M_2) / (\dot{M}_1 / M_1) > 1$ in our simulations is an important result, as it provides a mechanism for the distribution of mass-ratios for massive BH binaries to be biased toward equal mass.  

\begin{figure*}
\epsscale{0.5}
\plotone{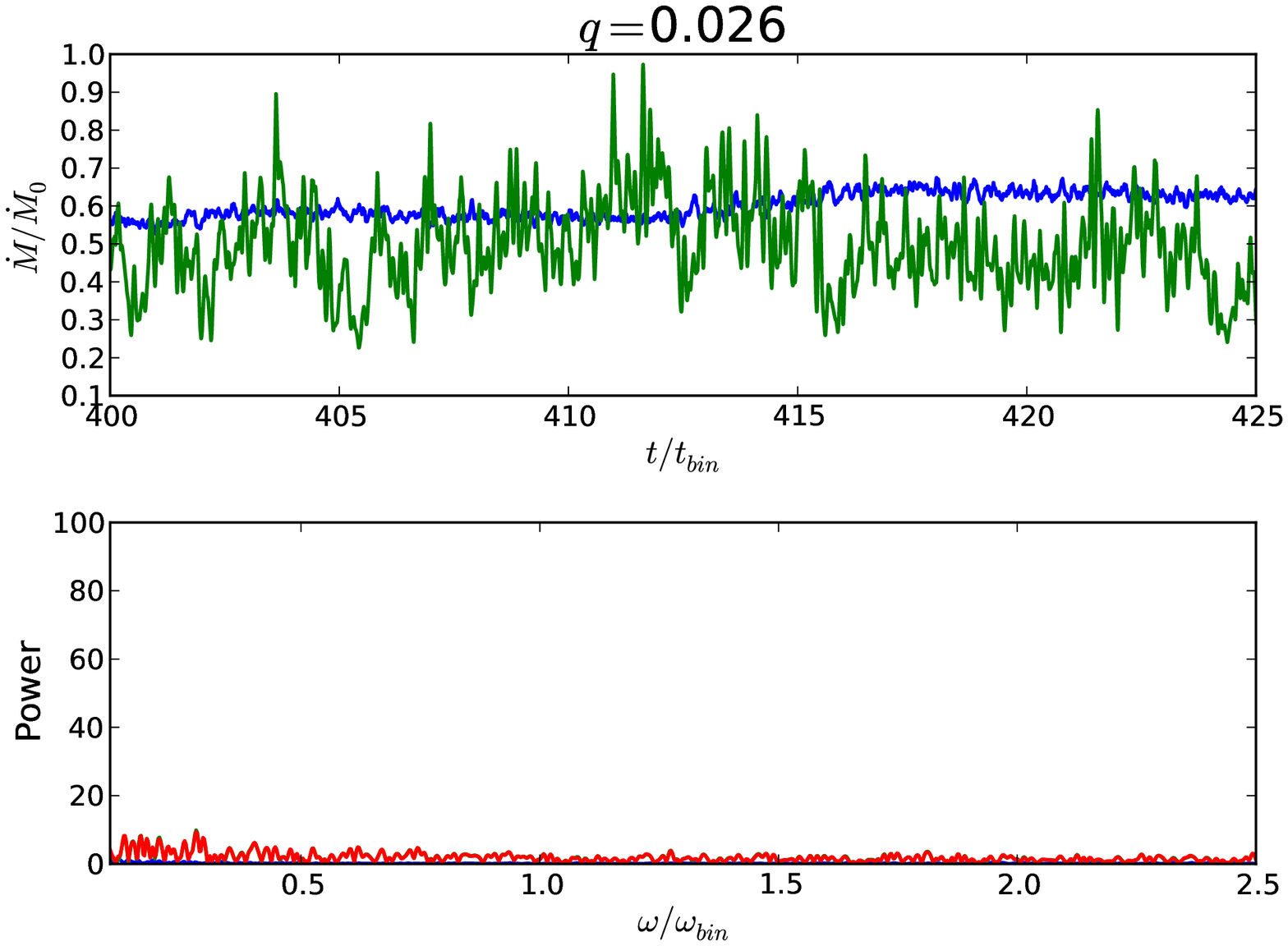}
\plotone{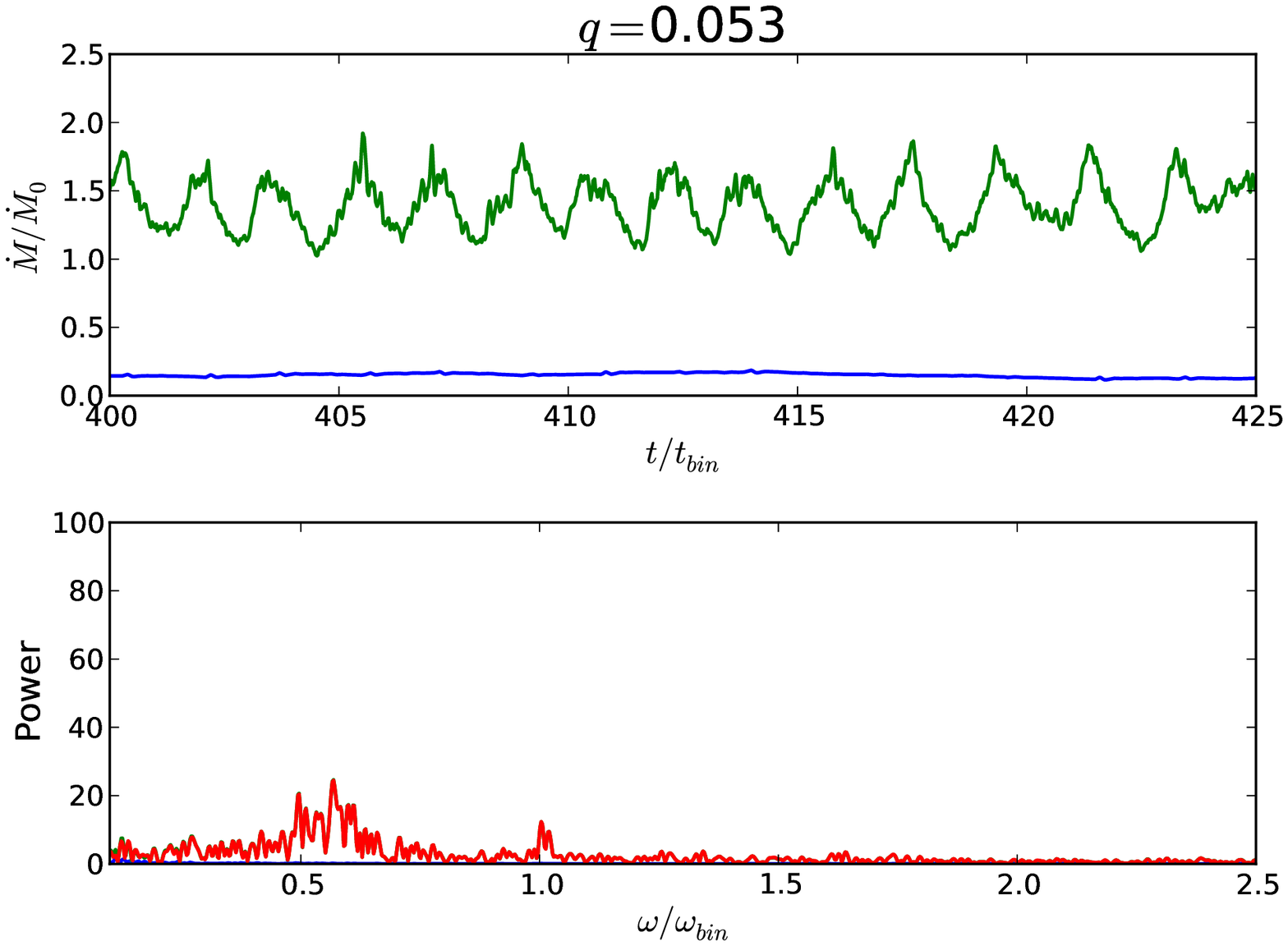}
\plotone{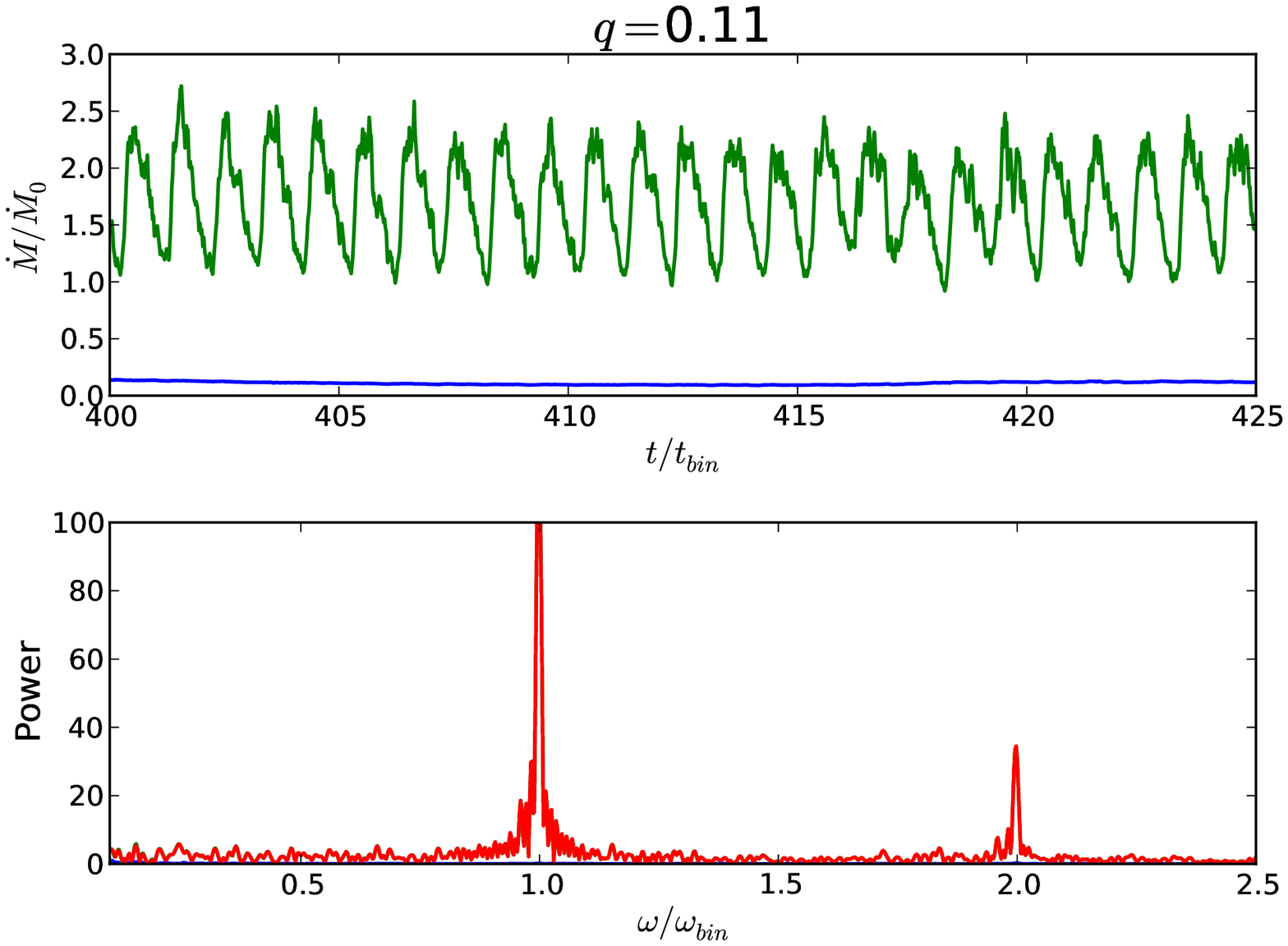}
\plotone{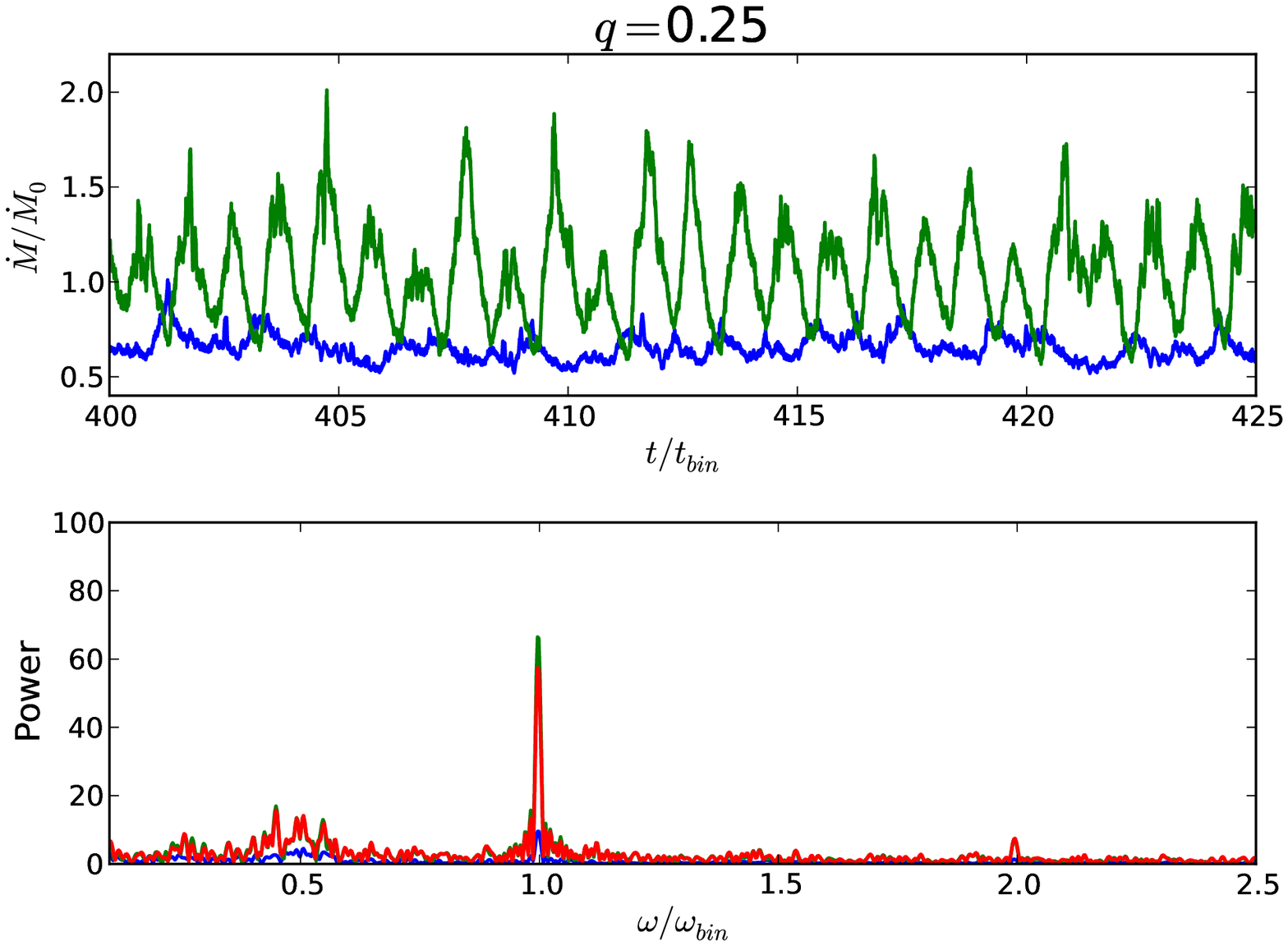}
\plotone{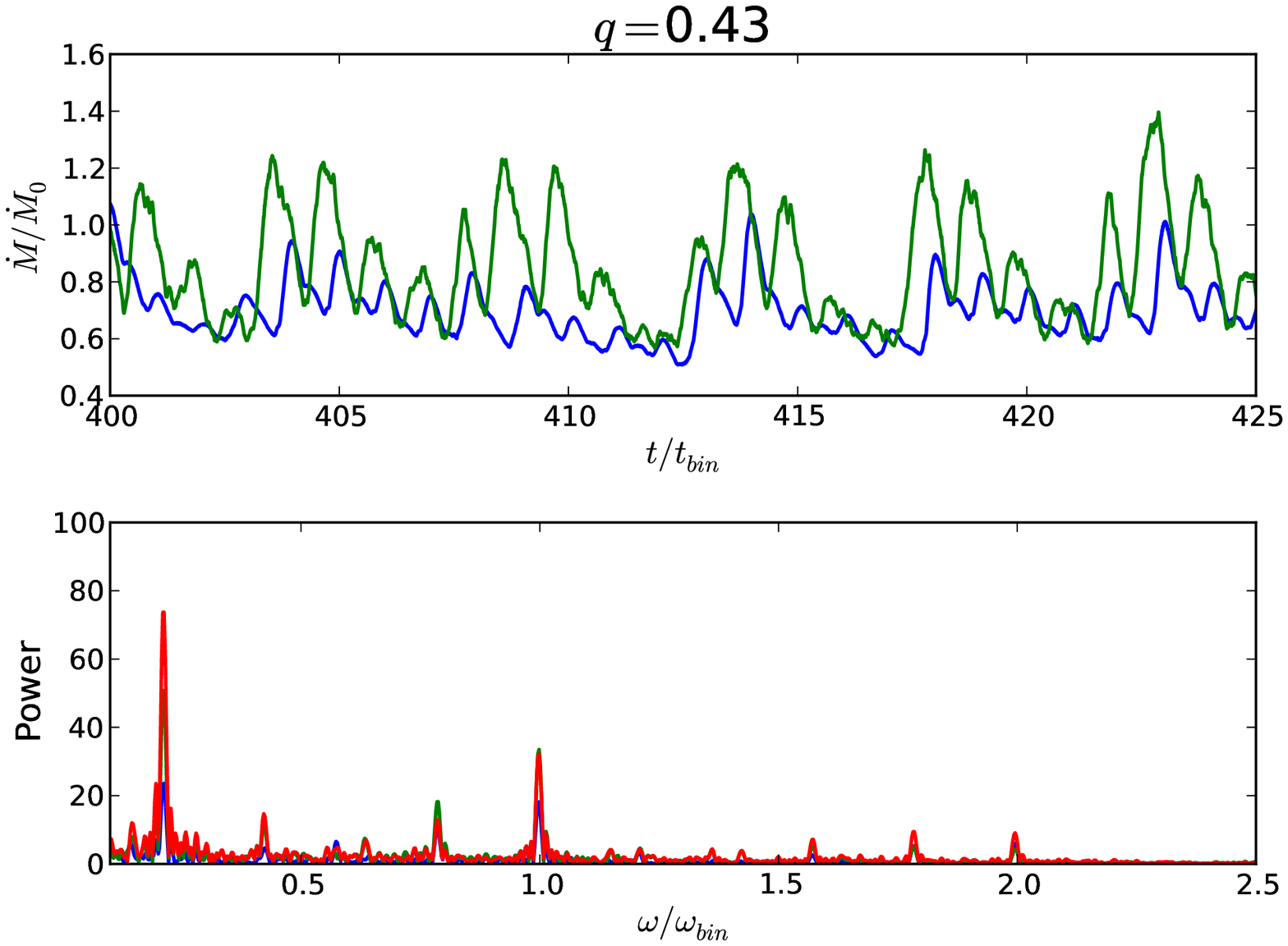}
\plotone{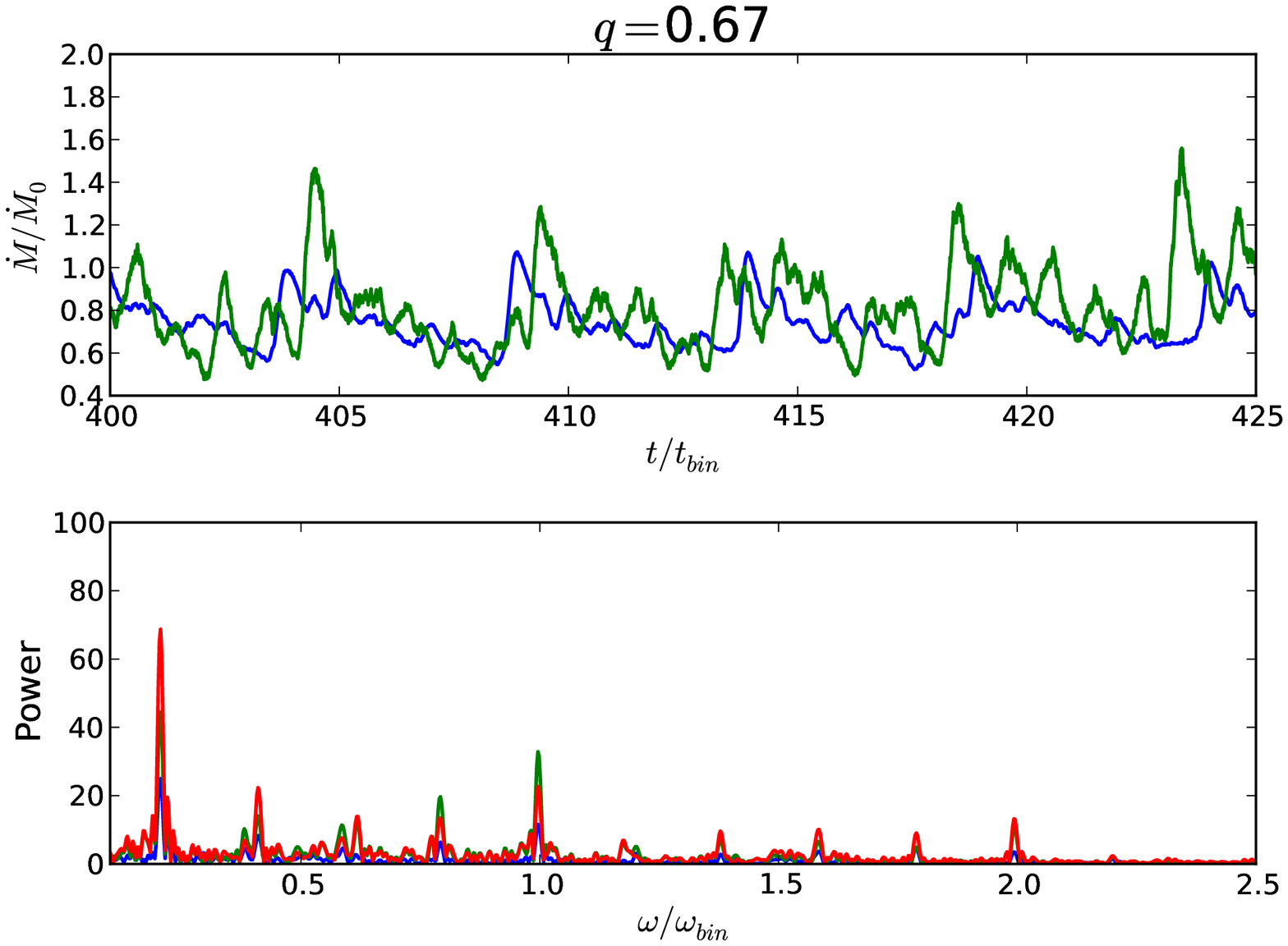}
\plotone{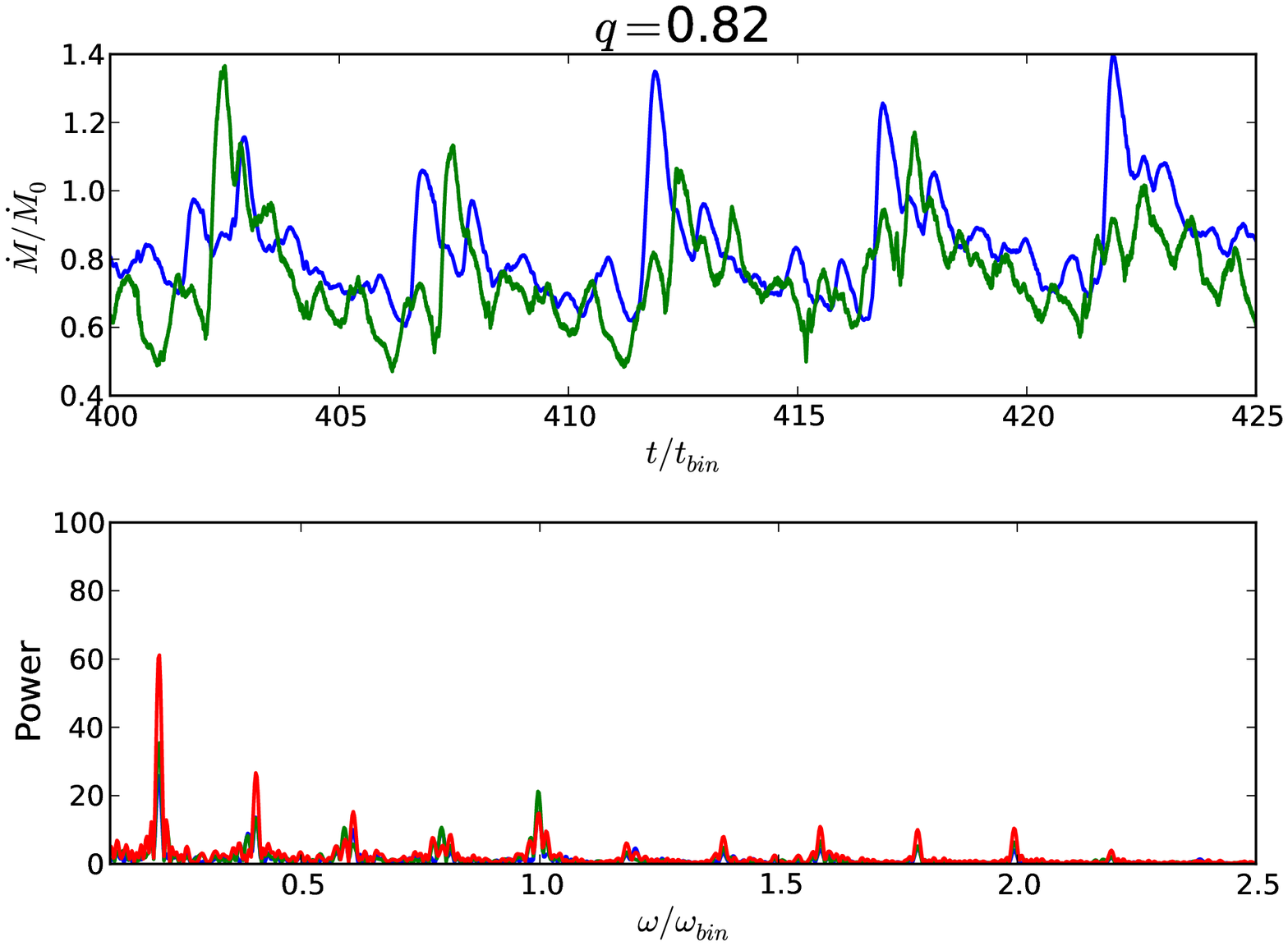}
\plotone{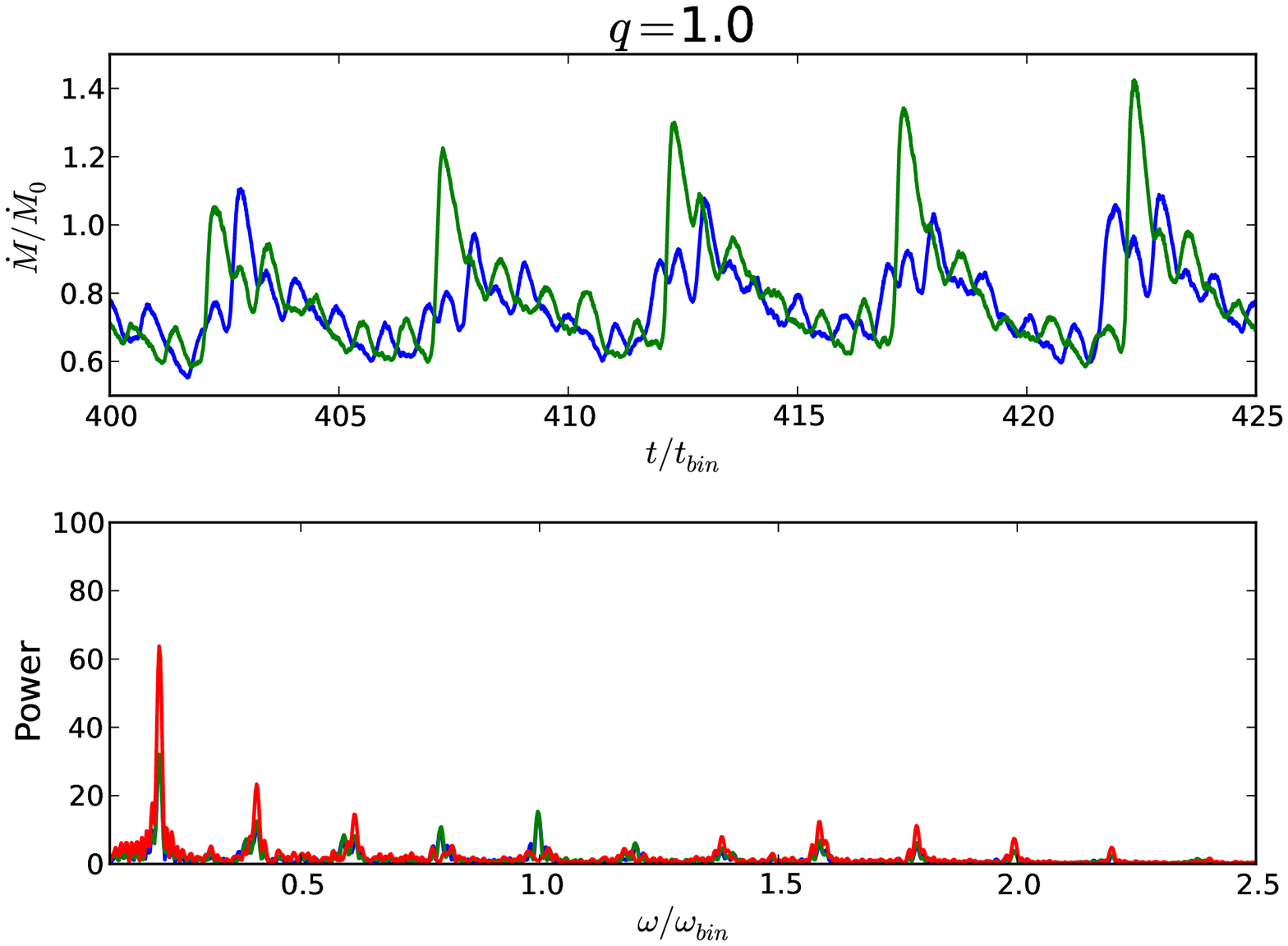}
\caption{The time variable accretion rates onto each BH, measured using Eq.~\ref{eq:tvis} and Eq.~\ref{eq:rhosink} (top of each pair of panels) and the corresponding Lomb-Scargle periodogram (bottom of each pair) computed over $\sim 100$ orbits. From left to right and top to bottom, panels correspond to mass ratios $q=0.026$, $0.053$, $0.11$, $0.25$, $0.43$, $0.67$, $0.82$, and $1.0$. In each panel, green lines correspond to the primary, blue lines correspond to the secondary, and red corresponds to the total accretion rate.}
\label{fig:periodogram}
\end{figure*}

It should be noted that while our smallest mass ratio, $q=0.026$, corresponds to a binary with increasing $q$, the accretion rate onto the primary actually marginally exceeds that of the secondary. In this case, the accretion flow has taken on a qualitatively different form, in which the accretion flow resembles an annular gap with a large circumprimary disk (see Figure~\ref{fig:dens2d}). Of course, this is the expected configuration for a binary with sufficiently small mass-ratio, and has been well studied in the context of protoplanetary disks \citep{lin93}. We find that in this case the cavity is no longer completely cleared, and significant gas is able to flow past the secondary onto a dense circumprimary disk, and the accretion rate onto the primary has risen relative to the $q=0.05$ case, and is slightly larger than that of the primary (see first panel of Figure~\ref{fig:periodogram}). This likely marks the beginning of a threshold in mass ratio $q_{crit}$ below which the torque from the secondary is no longer able to effectively clear an entire hollow central cavity and the morphology becomes more similar to that of a disk with an annular gap. Extrapolating the results of Figure~\ref{fig:mdot_ratio} to smaller mass ratios, $(\dot{M}_2 / M_2)/(\dot{M}_1 / M_1)$ may quickly drop below $1$, although more simulations are required to verify this prediction.
The value of $q_{crit}$ is likely highly dependent on the disk thickness as it depends on gas flowing past the secondary without becoming gravitationally bound. Consequently, our value of $q_{crit}\sim0.026$ should not be interpreted as a robust result applicable to all circumbinary disks. Nevertheless, this transition is important, because it may separate binaries into two categories: 1) large mass ratio binaries with $q>q_{crit}$ which are being further driven toward equal mass by enhanced accretion onto the secondary, and 2) small mass ratio binaries with $q<q_{crit}$ which may be further decreasing in mass ratio. This may lead to a bimodal binary mass-ratio distribution, with separate peaks at $q=1$ and below $q_{crit}$.

\begin{figure}
\epsscale{0.9}
\plotone{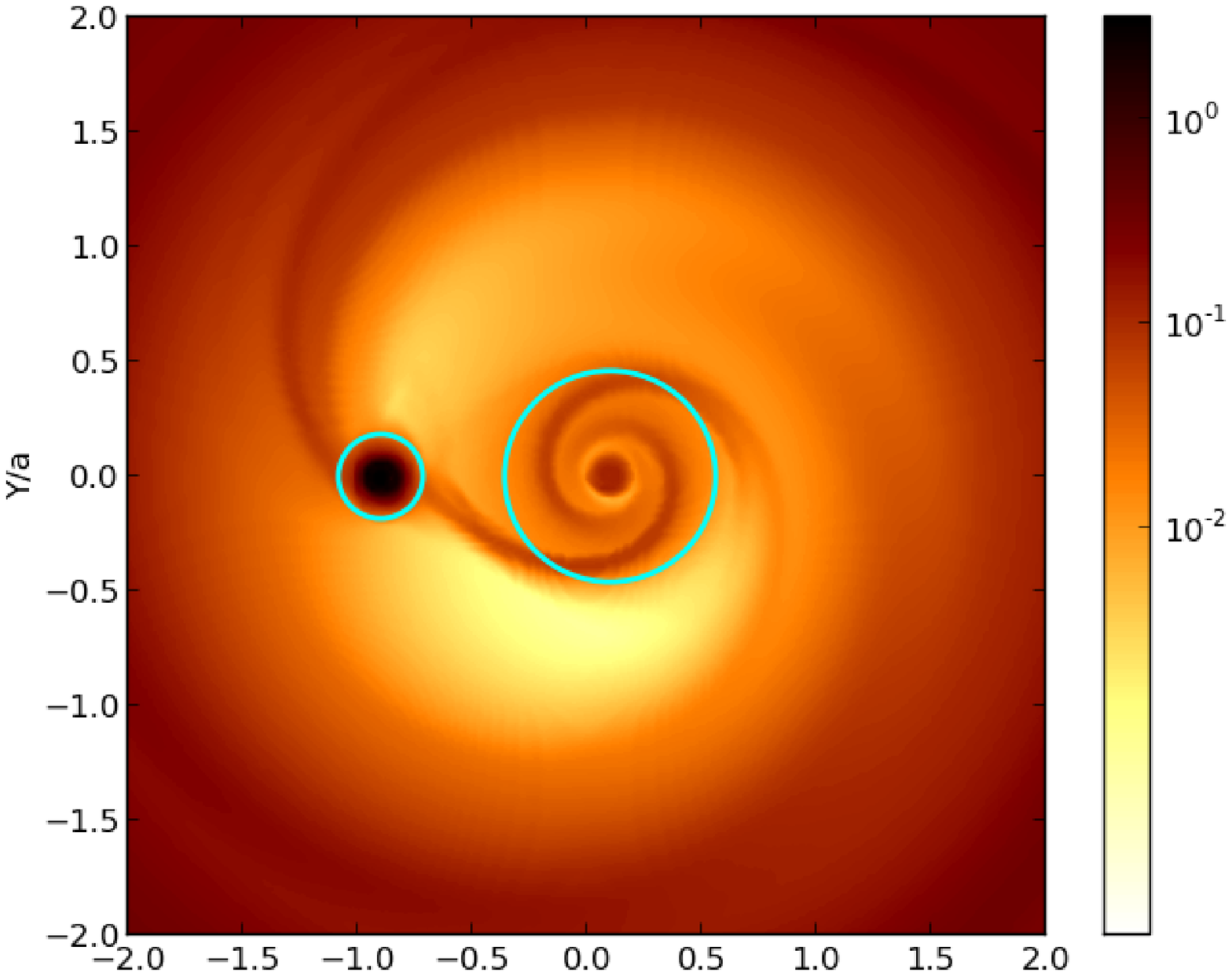}
\plotone{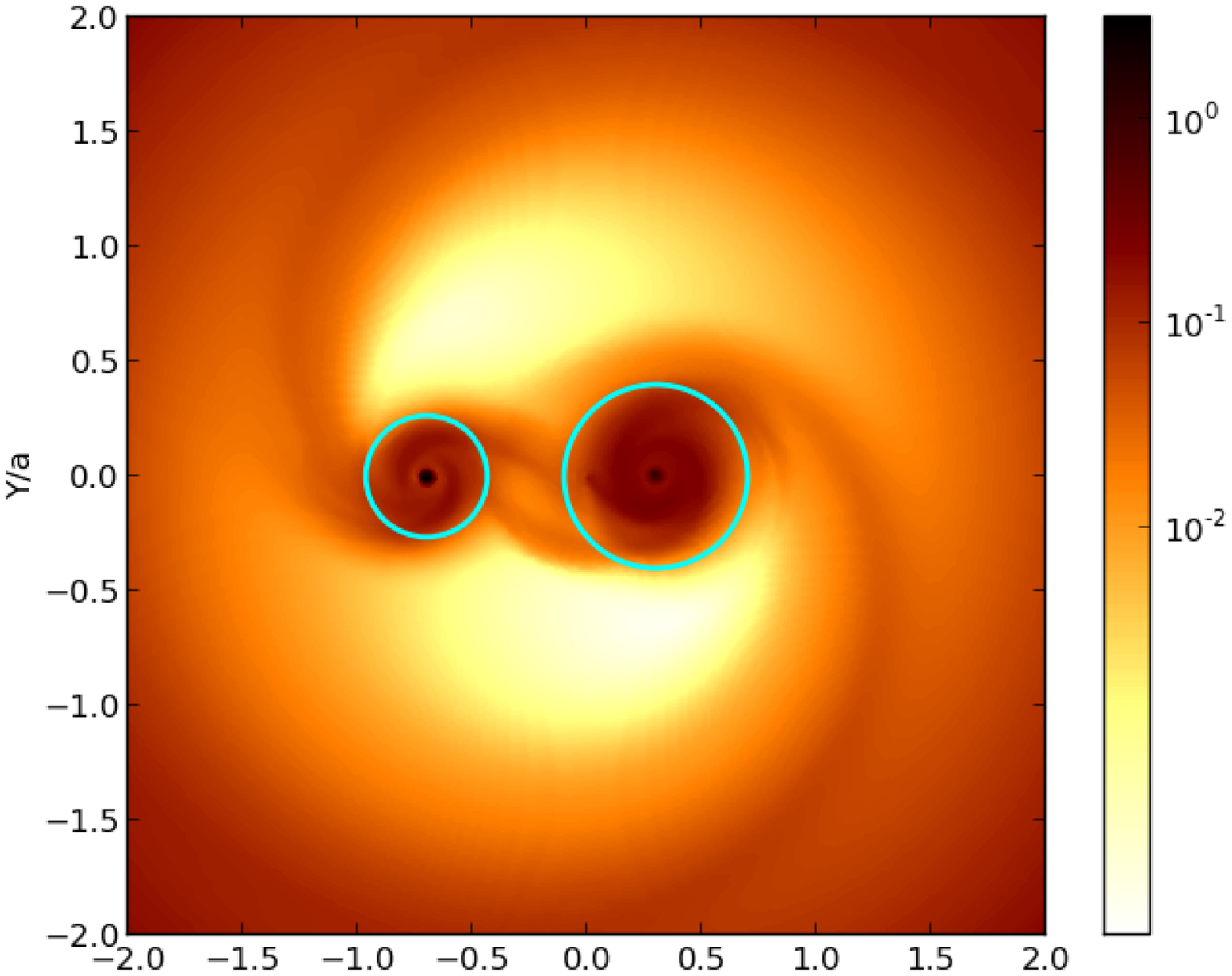}
\caption{Time-averaged surface density $\Sigma / \Sigma_0$. Cyan circles indicate the predictions of \citet{artymowicz94} for the size of the circumprimary and circumsecondary disks. Top figure is for a binary with $q=0.11$ ($\mu \equiv q/(q+1) = 0.1$). Bottom figure is for a binary with $q=0.43$ ($\mu \equiv q/(q+1) = 0.3$) }
\label{fig:tavg_surf_dens}
\end{figure}

In spite of the time-variability of the ``mini-disks", we can compare the time-averaged density profiles with analytic estimates of disk sizes based on perterbative calculations of disk truncation due to Lindblad resonances \citep{artymowicz94}. In Figure~\ref{fig:tavg_surf_dens}, we plot the time-averaged surface density profiles of the inner cavity region for the $q = 0.11$ and $q = 0.43$ cases (note that these are referred to as $\mu \equiv q/(q+1) = 0.1$ and $\mu \equiv q/(q+1) = 0.3$ in \citealt{artymowicz94}). We mark the analytic estimate of circumprimary and circumsecondary disk radii in cyan. In both cases, we find agreement with the analytic estimates, although we find that the circumprimary disk in the $q=0.11$ case is strongly perturbed by the nonaxisymmetric accretion streams. 

In Figure~\ref{fig:periodogram}, we plot the accretion rate onto each BH vs. time, as well as the Lomb-Scargle periodogram of the accretion rate for each mass ratio. We find many features in common with previous simulations in which the inner cavity at $r=a$ is excised from the grid and the total accretion rate is calculated through the inner boundary \citep{macfadyen08,dorazio13}. For example, we find that significant periodicity in the accretion rate emerges for mass ratios $q \gtrsim 0.1$, with peaks in the periodogram at $\omega /\omega_{bin}=1$ and $\omega / \omega_{bin}=2$, consistent with the findings of \citet{dorazio13} and \citet{roedig11}. These peaks are straightforwardly interpreted as arising from the passage of each BH near the overdense lump which forms at the edge of the eccentric cavity and subsequent stripping of material and replenishment of accretion streams. For $q<0.1$, these peaks are suppressed, as the binary torques are unable to excite eccentricity in the cavity and no noticeable overdensity is present. For $q=1.0$, the peak at $\omega / \omega_{bin}=2$ is present in the total accretion rate, but the peak at $\omega / \omega_{bin}=1$ is present only in each individual BH accretion rate. The combined accretion rate does not display this mode, as it is cancelled by the symmetry of this case. 

One notable feature of our simulations is the strength of the low frequency mode at $\omega_0 /\omega_{bin}\approx 0.2$. For each case in which $q \gtrsim 0.43$, we find that this is the dominant component in the periodogram. We note that a similar feature is also seen in MM08 and in the $\alpha=0.1$, $q=1$ case of \citet{dorazio13}. This low frequency component has been interpreted by MM08 as the orbital frequency of the overdense lump at the inner edge of the cavity. We also note that similar overdensities of gas near the inner wall of the cavity have been discussed in 3D MHD simulations by \citep{noble12,shi12}. Thus, any observation of periodicity in quasar emission is not likely to correspond directly to an associated binary period, but rather the orbital period of an overdense lump in the inner circumbinary accretion disk. In addition to the binary frequency and the lump frequency, we also find power in the associated harmonics. As in MM08, we find that for $q \gtrsim 0.43$, the total accretion rate can be decomposed into evenly spaced frequencies $\omega = \omega_0 K$, where $K = 1,2,3...$.

\begin{figure}
\epsscale{1.0}
\plotone{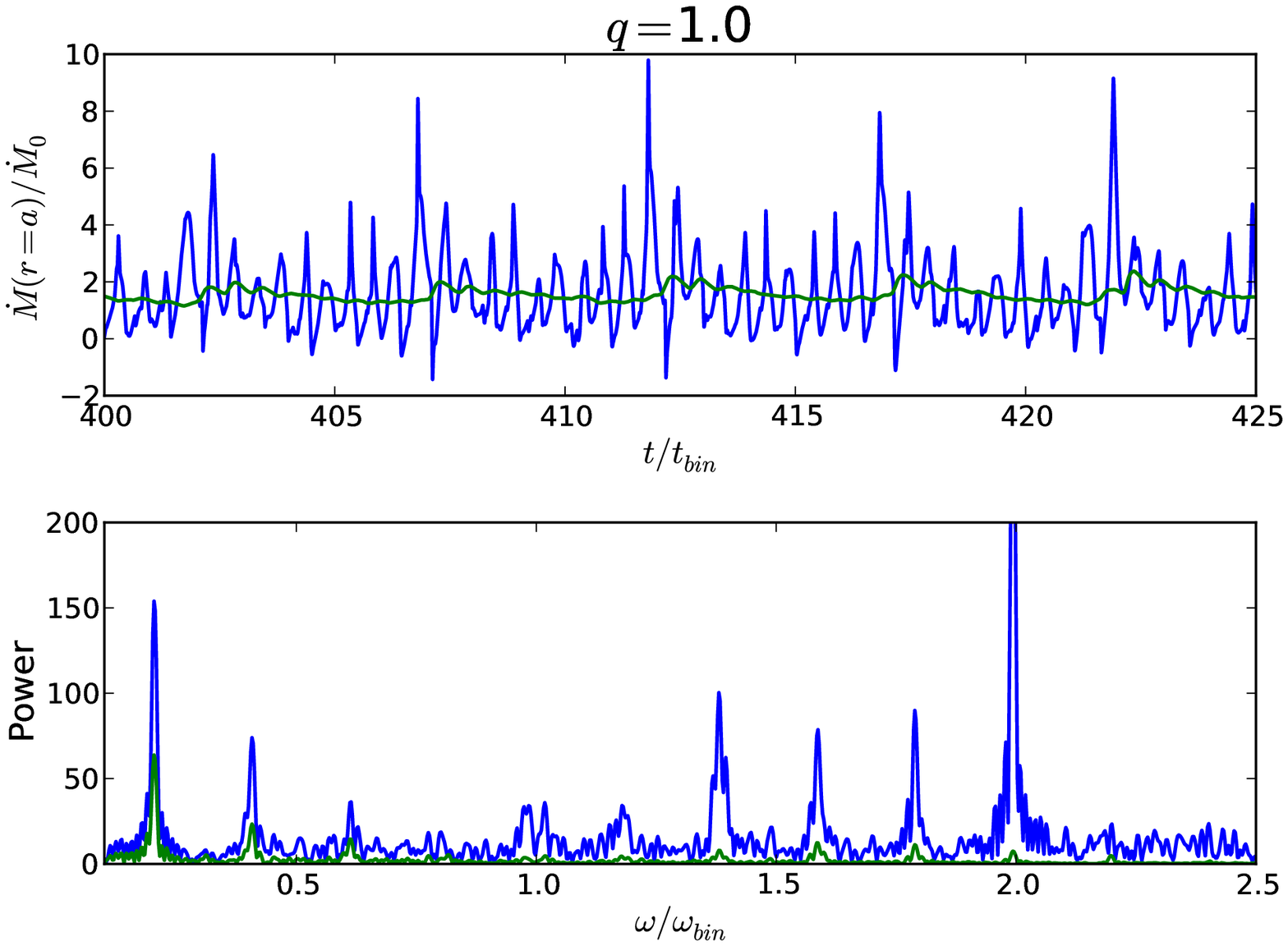}
\caption{The total accretion rate $\dot{M}$ through both black holes measured using two different methods. In the top panel in blue, we plot $\dot{M}$ computed by measuring the flux of rest mass through a surface at $r=a$. In the top panel in green, we plot $\dot{M}$ using Eq.~\ref{eq:tvis} and Eq.~\ref{eq:rhosink}. In the bottom panel, we plot the corresponding periodograms. We see that measurements of $\dot{M}$ that are performed at $r=a$ can overestimate the variability in the accretion rate. }
\label{fig:mdot_comparison}
\end{figure}

While the results of our simulations share some qualitative similarities with prior work, as described above, we find important differences. As our simulations do not excise the inner cavity, we are able to track the flow of the gas onto each minidisk, and track the accretion which takes place over a viscous timescale according to Eq.~\ref{eq:rhosink}. This is quite different from the accretion rate diagnostics employed in calculations which impose an inner boundary, in which $\dot{M}$ is directly linked to the flux of rest mass through the inner surface at $r=a$. We find that that the minidisks act as buffers, as gas accumulates in each minidisk and drains on a viscous timescale, rather than disappearing instantly upon entering the cavity. This smooths out the accretion rate and reduces the power in the periodogram at $\omega / \omega_{bin}=2$. This is illustrated in Figure~\ref{fig:mdot_comparison}, in which we plot the accretion rate measured according to Eq.~\ref{eq:rhosink}, and the flux of rest mass through a surface at $r=a$, for comparison. We find that although the two methods agree on the average accretion rate $\langle \dot{M} \rangle / \dot{M}_0 = 1.6$, the latter dramaticaly overestimates the variability. This highlights the importance of the dynamics inside the cavity in determining time-dependent binary accretion signatures. We speculate that our inclusion of the inner cavity may also explain why our measured accretion rate of $\dot{M} / \dot{M}_0 = 1.55$ for the $q=1$ case exceeds the corresponding measurement of $1.015$ reported in Table~2 of \citet{dorazio13}. By including the inner cavity in our simulations, we allow fluid elements on eccentric orbits to approach the binary more closely and to be ejected toward the cavity wall with greater energy. This is expected to create a more eccentric cavity, which in turn causes the lump to approach the secondary more closely at periastron, thus allowing more gas to be stripped.

\begin{figure}
\epsscale{1.0}
\plotone{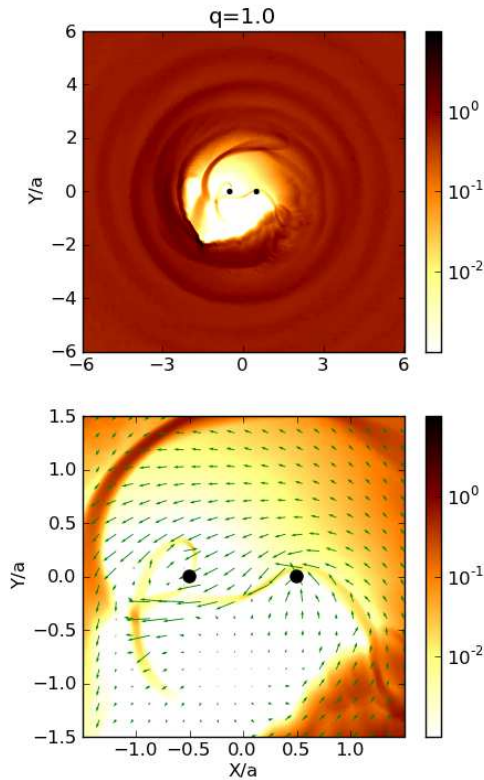}
\caption{Snapshots of surface density $\Sigma$ during quasi-steady state as in Figure~\ref{fig:dens2d}, except that the simulation was run with $t_{vis}$ shortened by a factor of $100$, restarting the $q=1$ run at $t=370 t_{bin}$ and running for another $100$ orbits. By shortening $t_{vis}$ so that it is less than $t_{bin}$, we find that ``mini-disks" are no longer able to form. We have added black circles to represent the locations of each BH.}
\label{fig:density2d_small_tvis}
\end{figure}

\begin{figure}
\epsscale{1.0}
\plotone{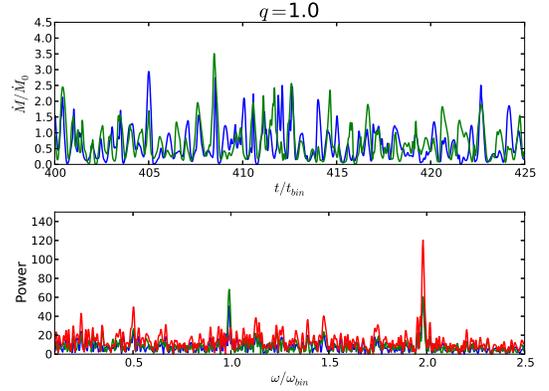}
\caption{The time variable accretion rates and periodograms as in Figure~\ref{fig:periodogram}, except that the simulation was run with $t_{vis}$ shortened by a factor of $100$, restarting the $q=1$ run at $t=370 t_{bin}$ and running for another $\sim 100$ orbits. In the absense of any ``mini-disks" to act as buffers, absorbing infalling gas and delaying accretion, the accretion shows much greater variability, and is dominated by the $\omega / \omega_{bin} = 2$ mode.}
\label{fig:periodogram_small_tvis}
\end{figure}

We have demonstrated that the periodicity in the measured accretion rate is strongly influenced by the presence of mini-disks inside the cavity. The size and variability of each mini-disk is, in turn, influenced by the temperature and viscosity inside the cavity, which determine the local accretion timescale. While our choices of viscosity and temperature profiles are physically motivated, they are subject to some uncertainty, and it is worthwhile to examine how the accretion flow inside the cavity changes when $t_{vis}$ is altered. To this end, we have restarted the $q=0$ simulation from a checkpoint at $t/t_{bin}=370$ and evolved for another $\sim 100$ orbits with the viscous timescale $t_{vis}$ from Eq.~\ref{eq:tvis} shortened by a factor of $100$. From Eq.~\ref{eq:tvis_md}, we see that this shortens the accretion timescale to less than the orbital timescale. Thus, any gas which is stripped during a passage with the lump should accrete within one orbit before having time to form a minidisk. Indeed, we see in Figure~\ref{fig:density2d_small_tvis} a 2D snapshot of the surface density in which the minidisks are completely absent. This has the expected effect on the measured accretion rate, as the buffering effect of the mini-disks is lost and the accretion variability is increased, as shown in Figure~\ref{fig:periodogram_small_tvis}.  In this case, the frequency at $\omega / \omega_{bin} = 2$ dominates, corresponding to the passage of the BHs past the lump which occurs twice per orbit. We emphasize that although the shortening of $t_{vis}$ here is artificial, it is not implausible that such short timescales could be realized in nature. As $t_{vis} \sim (h/r)^{-2} \alpha^{-1}$, significant heating of the minidisk, or a mechanism for generating a large effective viscosity inside the cavity may significantly shorten $t_{vis}$. This underscores the need for future work to include more sophisticated treatments of the heating and cooling processes that determine the temperature inside the cavity, as well as precise treatments of viscosity that are motivated by studies of MHD turbulence.

For each mass ratio, we have computed the torque density profile following the definition provided in MM08,
\begin{equation}
\frac{dT}{dr} = \left\langle \frac{1}{2 \pi} \int_0^{2\pi} \Sigma(r,\phi)\frac{d\Phi}{d\phi}(r,\phi) r d\phi \right\rangle \ .
\end{equation}
and find that our results match those of MM08 very well for $q \gtrsim 0.43$. For lower mass ratios, we find that the peaks are shifted inward. This is consistent with the fact that smaller mass ratios lead to smaller cavities (See Figure~\ref{fig:vs_q}). This is not surprising, as torques beyond $r/a \gtrsim 2$ should be much weaker for binaries with small $q$, and we expect the disk to be less distorted by a weaker perturber.
\begin{figure}
\epsscale{1.0}
\plotone{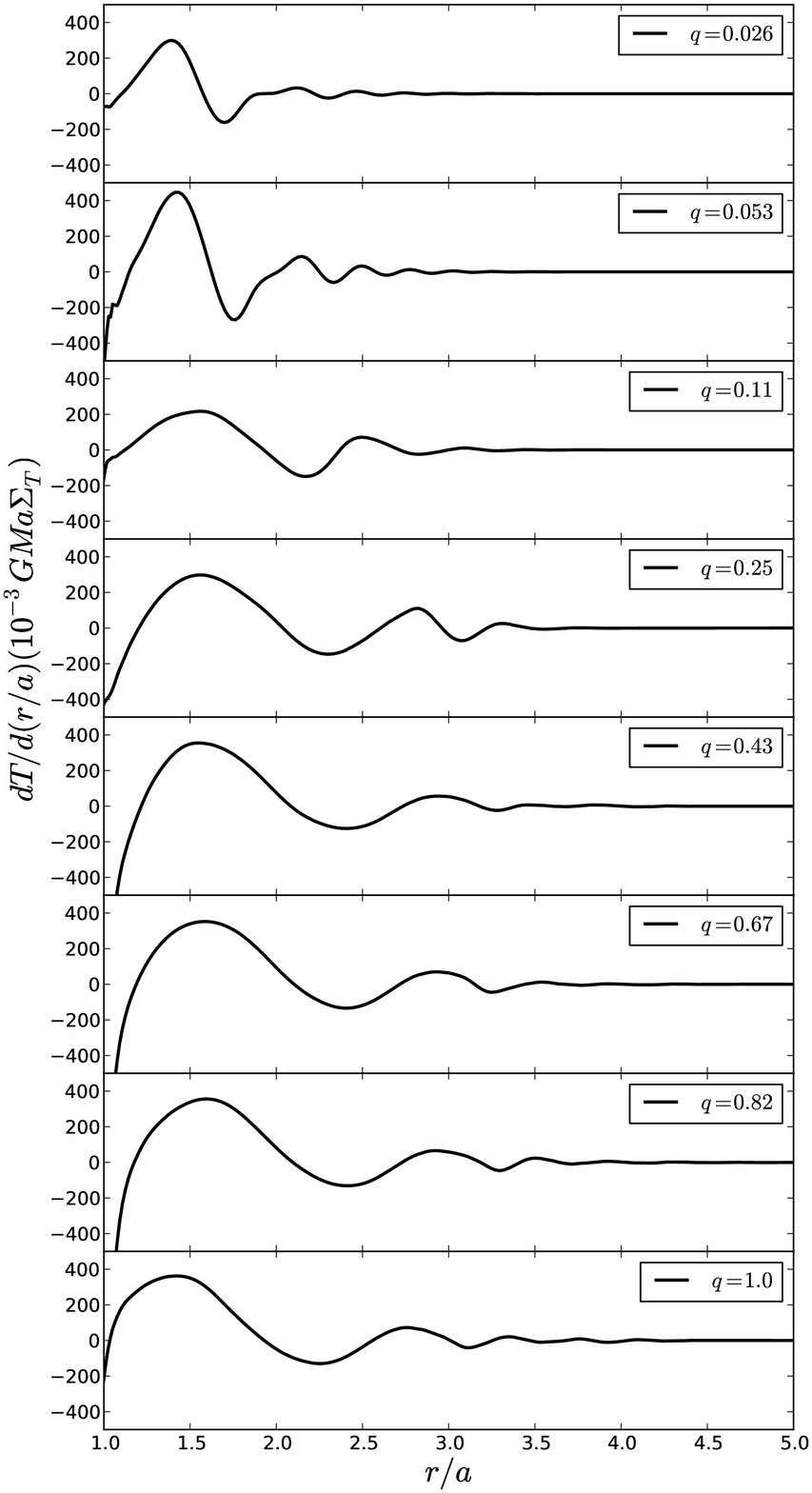}
\caption{Time-averaged torque density $dT/dr$ as a function of radius for various binary mass ratios. Each torque density is normalized by $\Sigma_T$, which is defined to be the time-averaged, angle-averaged surface density at the location of the first peak in $dT/dr$. Time average is taken over $446 < t_{bin}/M < 462$.}
\end{figure}
\label{fig:torque}

\section{summary and future work}
\label{sec:conclusions}
\subsection{Summary}
We have performed the first hydrodynamic simulations of thin, circumbinary accretion disks using a moving-mesh, finite volume code with the BHs present on the grid. Our disks are locally isothermal, corotating in the plane of the binary, and employ an $\alpha-$law viscosity prescription. Our binaries are held at fixed separation in a circular orbit, and we simulate a range of mass ratios $0.026 \le q \le 1.0$. In each case, we ensure that we evolve for more than one viscous time at the inner edge of the circumbinary cavity, so that the inner accretion region has fully relaxed to a quasi-steady state.

The important results of this work are the following:
\begin{enumerate}
\item We confirm the finding of prior work \citep{macfadyen08,dorazio13} that the accretion rate onto the BHs is not significantly reduced by the presence of a binary, when compared to the accretion rate onto a single BH of the same mass. This is the case in spite of the fact that much of the inner cavity is cleared of gas by the action of the binary torques, and is due to the effectiveness of the narrow accretion stream in delivering gas from the circumbinary disk inner edge to the individual black holes.
\item For each mass ratio considered, we find that ``mini-disks" surrounding each BH are formed. In each case, the mini-disks are persistent, as their accretion timescale greatly exceeds the binary orbital timescale. However further work is required to determine whether or not this remains true when mini-disks are allowed to become hotter. For the $q=0.11$, and $q=0.43$ cases, we confirm that the size of these mini-disks is in rough agreement with the predictions of \citet{artymowicz94}.
\item We find that significant periodicity in the accretion rates emerges only for $q \gtrsim 0.1$. At these mass ratios, the binary torques are strong enough to excite eccentricity in the inner cavity and create an overdense lump, whose interaction with the passing BHs leads to periodicity in the accretion rate. We find a strong peak in the periodograms for these cases correponding to the orbital frequency of the lump, with many associated harmonics for the $q\gtrsim 0.43$ cases. This periodicity may consititute a unique observational signature of SMBH binaries.
\item We find that for each case considered here, the accretion rate onto the secondary is sufficiently large relative to that of the primary so that the mass ratio $q$ is increasing. Similar results have been found previously in SPH calculations \citep{hayasaki07,cuadra09,roedig11,roedig12}. As SMBHs are expected to gain a significant fraction of their mass through gas accretion \citep{soltan82,yu02,elvis02}, this suggests a mechanism which may bias the distribution of binaries near merger toward higher mass ratios.
\end{enumerate}

\subsection{Future Work}
In future work, we intend to relax several assumptions, while expanding the parameter space of our simulations. In this paper we have assumed a locally isothermal equation of state with the temperature set such that $h/r \approx 0.1$ for both the circumbinary disk as well as each ``mini-disk". This assumption should ideally be relaxed by incorporating a more sophisticated treatment of the temperature, so that the mini-disks are allowed to heat or cool independently. This may have important consequences for the periodicity of the accretion, as the accretion timescale is sensitive to the temperature. If the accretion timescale becomes smaller than the binary orbital timescale, we may find that the accretion occurs in periodic outbursts, as discussed in \citep{tanaka13}. We also expect that realistic circumbinary disks may have a range of thicknesses, and we intend to investigate other choices for $h/r$.

In this work we have worked in the limit of $M_{disk} \ll M$, and thus kept the binary on a fixed circular orbit. In future work, this assumption should be relaxed in order to determine how binary eccentricity can be influenced by the eccentricity present in the inner disk. If significant eccentricity is imparted to the binary, it may have observational consequences for gravitational wave observations.

By working in the limit of $M_{disk} \ll M$, we have also neglected the shrinkage of the binary orbit due to interactions in the disk which can take place over long timescales \citep{kocsis12a,kocsis12b}. In future work, simulations should be done to characterize the coupled disk+binary evolution.

In each of our simulations, we have assumed that the orientation of the disk is aligned with that of the binary. Recent work done with SPH simulations of misaligned (or retrograde) disks indicate that dramatic differences may emerge when the binary is inclined with respect to the disk \citep{nixon13,roedig13}, but have yet to be compared with results of finite-volume simulations. 

Finally, we have ignored BH spin in these calculations. In future work, including spin effects may modify the accretion flow onto individual BHs, and monitoring the change in the spin of each BH due to gas accretion may provide insight into models of binary BH spin distributions.

In summary, we have found that accretion flows surrounding BH binaries exhibit a host of complex features, including uneven BH growth rates which depend on mass-ratio, potentially leading to biases in the binary BH mass-ratio distribution, as well as distinctive periodicity in BH accretion rates, which may provide a method of probing binary parameters through electromagnetic observations. Future simulations that incorporate the ingredients listed above will shed further light on these issues.

\section{acknowledgements}
We would like to thank Daniel D'Orazio, Michael Kesden, Takamitsu Tanaka, and Aleksey Generozov for useful discussions. We acknowledge support from NASA grant NNX11AE05G (to ZH and AM) and by the NSF through grant AST-1009863 (to AM). Resources supporting this work were provided by the NASA High-End Computing (HEC) Program through the NASA Advanced Supercomputing (NAS) Division at Ames Research Center.

\bibliographystyle{apj}
\bibliography{refs}

\end{document}